\documentclass[12pt]{iopart}
\usepackage{iopams}
\usepackage{epsfig}
\newlength{\bredde}
\def\slash#1{\settowidth{\bredde}{$#1$}\ifmmode\,\raisebox{.15ex}{/}
\hspace*{-\bredde} #1\else$\,\raisebox{.15ex}{/}\hspace*{-\bredde} #1$\fi}

\newcommand{\be}{\begin{equation}}
\newcommand{\ee}{\end{equation}}
\newcommand{\bea}{\begin{eqnarray}}
\newcommand{\eea}{\end{eqnarray}}
\newcommand{\nn}{\nonumber}
\newcommand{\sect}[1]{\setcounter{equation}{0}\section{#1}}

\def\erfc{{\mbox{erfc}}}
\def\sgn{{\mbox{sgn}}}
\def\Tr{{\mbox{Tr}}}
\def\Pf{{\mbox{Pf}}}
\def\re{{\Re\mbox{e}}}
\def\im{{\Im\mbox{m}}}

\def\La{\Lambda}
\def\la{\lambda}
\def\al{\alpha}
\def\e{{\mbox{e}}}

\newcommand{\mcA}{\mathcal{A}}
\newcommand{\mcD}{\mathcal{D}}
\newcommand{\mcF}{\mathcal{F}}
\newcommand{\mcK}{\mathcal{K}}

\begin{document}
\title[Chiral real asymmetric 2-matrix model]{The chiral Gaussian
two-matrix  ensemble of real asymmetric matrices}

\author{G. Akemann$^1$, M. J. Phillips$^1$, and H.-J. Sommers$^2$}

\address{\it
$^1$Department of Mathematical Sciences \& BURSt Research Centre,\\
\ Brunel University West London, Uxbridge UB8 3PH, United Kingdom\\
$^2$Fakult\"at f\"ur 
Physik, Universit\"{a}t Duisburg-Essen, 47048 Duisburg, Germany
}
\begin{abstract}
We solve a family of Gaussian two-matrix models with rectangular
$N\times(N+\nu)$ matrices,
having real asymmetric matrix elements and depending on a non-Hermiticity
parameter $\mu$. Our model can be thought of as the chiral extension of the
real Ginibre ensemble,
relevant for Dirac operators in the same symmetry class. 
It has the property that its eigenvalues are either
real, purely imaginary, or come in complex conjugate eigenvalue pairs.
The eigenvalue joint probability distribution for our
model is explicitly computed,
leading to a non-Gaussian distribution including $K$-Bessel functions.
All $n$-point density
correlation functions are expressed for finite $N$ in terms of a Pfaffian
form. This contains a kernel involving Laguerre polynomials
in the complex plane as a building block which was previously computed by the
authors.
This kernel can be expressed in terms of the kernel for \textit{complex}
non-Hermitian 
matrices, generalising the known relation among ensembles of Hermitian random
matrices.
Compact expressions are given for the density
at finite $N$ as an example, as well as its microscopic large-$N$
limits at the origin for fixed $\nu$ at strong and weak non-Hermiticity.
\end{abstract}



\sect{Introduction}

Non-Hermitian Random Matrix Theory (RMT) introduced by Ginibre
\cite{Ginibre65} is
almost as old as its Hermitian counterpart. At first it was seen as an
academic exercise to drop the Hermiticity constraint and thus to allow for
complex eigenvalues. However, in the past two decades we have seen many
applications of such RMTs featuring complex eigenvalues precisely for physical
reasons, and we refer to \cite{FS} for examples and references.
Because matrices with real data are often modelled by RMT one could view
the real Ginibre ensemble of asymmetric matrices as being the most interesting
non-Hermitian ensemble. Unfortunately it has also turned out to be
the most difficult one, as it took over 25
years to compute the joint distribution of its eigenvalues
\cite{Lehmann91,Edelman97},
being real or coming in
complex conjugate pairs.
The integrable structure and all
eigenvalue correlation functions were computed only very recently
for the real Ginibre ensemble
\cite{Akemann07,Sinclair06,Sommers2007,Forrester07,
Borodin07,Sommers2008,ForresterMays}.

Our motivation for generalising this model is as follows. In the 1990's
Verbaarschot proposed extending  the three classical (and Hermitian) ensembles
of Wigner and Dyson to so-called chiral RMT \cite{SV93},
in order to describe the low energy sector
of Quantum Chromodynamics (QCD) and related field theories.
These chiral ensembles are also known as Wishart
or Laguerre ensembles.
Their non-Hermitian extensions \cite{Steph,HOV} were motivated by adding a
chemical potential for the quarks, which breaks the anti-Hermiticity of the
Dirac operator in field theory. It was observed numerically quite early
\cite{HOV}
that these chiral versions of the Ginibre ensembles have distinct
features, either attracting eigenvalues to the real and imaginary axes (real
matrices), repelling them (quaternion real matrices) or having no such symmetry
(complex matrices). Only later was it realised how to solve
these chiral non-Hermitian RMTs analytically, by using replicas
\cite{KJreplica} or by extending the initial
one-matrix model plus a constant symmetry-breaking term
\cite{Steph,HOV} to a two-matrix
model. This idea from Osborn \cite{Osborn} led to a complex eigenvalue model
that can be solved using
orthogonal polynomials in the complex plane \cite{A03}. The solution of the
two-matrix model
was then derived for complex
\cite{Osborn,AOSV} and quaternion real matrices \cite{AB}.
Our paper aims to solve the third and most difficult of such
non-Hermitian RMTs, a chiral two-matrix model of real asymmetric matrices
introduced in our previous work \cite{APSo}.
For more details on RMT applications to the QCD-like Dirac operator spectrum we
refer to \cite{A07mu}.

Many more non-Hermitian RMTs than just the three Ginibre ensembles and their
chiral (or Wishart/Laguerre)
counterparts exist \cite{Magnea08} and these are mostly unsolved to date.
Very recently another two-matrix model generalisation of the real Ginibre
ensembles was introduced and solved in \cite{FM09}. There the eigenvalue
correlations of the {\it ratio} of two quadratic matrices are sought, whereas
here we 
deal with the {\it product} of two rectangular matrices. Whilst the former case
leads to a Cauchy-type weight function, in our model we will obtain a weight of
Bessel-$K$ functions for the eigenvalues.
We hope that given the
plethora of RMT applications, our particular
model will find applications beyond the
field theory that it has been designed for.

The approach of solving our model is based on the variational method
detailed in
\cite{Sommers2007,Sommers2008}.
It follows its two main ideas: first to compute the
joint probability distribution function (jpdf) for general $N$ by reducing it
to $2\times2$ and $1\times1$ blocks. Because we are considering rectangular
matrices this is \textit{a priori} not guaranteed to work.
Second, we use the variational method \cite{Sommers2007,Sommers2008} in
combining all density correlations with $n$ points (being
real, purely imaginary or complex conjugates) into a single Pfaffian
form. This reduces the computation to determining its main building block, an
anti-symmetric kernel. Whilst it can be deduced from the spectral 1-point
density -- which was known for the real Ginibre ensemble \cite{EKS94}
-- we here exploit an idea from our previous publication
\cite{APSo}. There the kernel was determined by computing the expectation value
of two characteristic polynomials using Grassmannians. The same relation
between kernel and characteristic polynomials is known to hold
for the symmetry classes with complex \cite{AV} or quaternion real
matrices \cite{AB}, in fact for any class of non-Gaussian weight functions.

As a new result we can express our kernel for real asymmetric
matrices in terms of the kernel for complex non-Hermitian matrices. Such a
relation might have been expected to exist as it is known for Hermitian RMT
\cite{TW,F99}.

Other methods that have been applied successfully to the real Ginibre
ensemble such as the supersymmetric method \cite{Efetov97}, skew-orthogonal
polynomials \cite{Forrester07} 
or probabilistic methods \cite{Borodin07}
are very likely to be extendible to our two-matrix model as well.

The paper is organised as follows. In Section \ref{sum} we summarise
our main statements: the definition of the matrix model, its
jpdf in terms of the real, imaginary  and complex conjugate eigenvalue
pairs, and the solution for all density correlation functions as a Pfaffian of
a matrix-valued kernel. Examples are given for the simplest spectral
densities at finite $N$ and in the
microscopic large-$N$ limits for
strong and for weak non-Hermiticity at the origin.
These findings are then detailed
in Section \ref{jpdf} on the jpdf, where we separately treat $N=1,2$ and
general $N$. The spectral density correlations and their finite- and
large-$N$ results are derived and illustrated
in Section \ref{dense}. Our conclusions are presented in Section \ref{conc}.
Some technical details on the computation of the
Jacobian are collected in \ref{Jacobi}.

\sect{Summary of results}\label{sum}

\subsection{The model}

The chiral Gaussian ensemble of real asymmetric matrices
as introduced by the authors \cite{APSo} is given by a
two-matrix model
of rectangular matrices $P$ and $Q$ of sizes $N\times (N+\nu)$ with real
elements, without further
symmetry restriction. The partition function normalised to unity is given by
\be
{\cal Z}=\left(\frac{1}{\sqrt{2\pi}}\right)^{2N(N+\nu)}
\int_{\mathbb{R}^{N(N+\nu)}}\!\!\!\!\!\!\!\!dP
\int_{\mathbb{R}^{N(N+\nu)}}\!\!\!\!\!\!\!\! dQ
\ \exp\left[-\frac12\Tr(PP^T+QQ^T)\right]\ ,
\label{ZPQ}
\ee
where we integrate over all the independent, normally
distributed matrix elements of $P$ and $Q$.
We are interested in the eigenvalues of the matrix $\mcD$ of size
$2N+\nu$ squared
\be
\mcD\equiv \left(\begin{array}{cc}
0& P+\mu Q\\
P^T-\mu Q^T&0\\
\end{array}
\right)
\equiv
\left(\begin{array}{cc}
0& A\\
B^T&0\\
\end{array}
\right)
\ .
\label{Mdef}
\ee
Here $\mu\in(0,1]$ is the non-Hermiticity parameter, interpolating between the
  chGOE $(\lim\mu\to0)$ and maximal non-Hermiticity $(\mu=1)$. The analogous
  chiral Gaussian two-matrix models with unitary and symplectic symmetry were
  introduced in \cite{Osborn,AB} respectively.

In applications to field theory, $\mcD$ corresponds to the chiral Dirac
operator, and $\mu$ to the chemical potential\footnote{The Euclidian Dirac
  operator in field theory is actually anti-Hermitian for $\mu=0$, but we will
not use this convention here.}.
Typically, $N_f$ extra determinants of the type $\det[\mcD+mI_{2N+\nu}]$ are
inserted
into the partition function eq.\ (\ref{ZPQ}), where $m$ is the quark mass,
but we will restrict ourselves in this paper to
the case $N_f=0$; this is referred to as the quenched case.

For later convenience we give an equivalent form of eq.\ (\ref{ZPQ}),
by changing variables from
\be
P=\frac12(A+B)\ ,\ \ Q=\frac{1}{2\mu}(A-B)\ ,
\label{PQdef}
\ee
to matrices $A$ and $B$ defined in eq.\ (\ref{Mdef}):
\bea
\fl{\cal Z}&=&\left(\frac{1}{4\pi\mu}\right)^{N(N+\nu)}
\int_{\mathbb{R}^{N(N+\nu)}}\!\!\!\!\!\!\!\!dA
\int_{\mathbb{R}^{N(N+\nu)}}\!\!\!\!\!\!\!\!dB
\ \e^{-\frac12\eta_+\Tr(AA^T+BB^T)
+\frac12\eta_-\Tr(AB^T+BA^T)}\ ,
\label{ZAB}\\
\fl&&\mbox{with}\ \ \eta_\pm\ \equiv\ \frac{1\pm\mu^2}{4\mu^2}
\label{xedef}\ .
\eea
The two $\mu$-dependent combinations $\eta_\pm$ will
be used throughout the paper.

\subsection{Eigenvalue representation}

The eigenvalues $\La$ of the Dirac matrix $\mcD$ are determined from the
following equation\footnote{This follows from eq.\ (\ref{Mdef}) using the
  standard relation
$\det\left(
\begin{array}{cc}a&b\\c&d
\end{array}
\right)=\det(d)\,\det (a-bd^{-1}c)$ for square matrices $a$ and $d$, with
$d$ invertible.}:
\be
0=\det[\Lambda I_{2N+\nu}-\mcD]=\La^\nu\det[\La^2 I_N-AB^T]=
\La^\nu\prod_{j=1}^N(\La^2-\La_j^2)\ .
\label{evrel}
\ee
For this reason we will first compute the eigenvalue distribution of the
$N\times N$ Wishart-type combination of matrices
$C\equiv AB^T$. $C$ has real elements, and therefore its eigenvalues
$\La_j^2$ are real, or else come in complex conjugate pairs.
The matrix $\mcD$ itself has the following solutions: $\nu$ zero-eigenvalues
$\La=0$, and $2N$
eigenvalues coming in pairs $\La=\pm\La_j$.
Consequently the non-zero eigenvalues of $\mcD$ fall into three categories:
\begin{enumerate}

\item for $\La_j^2>0$: real pairs $\La=\pm\La_j\in\mathbb{R}$

\item for $\La_j^2<0$: purely imaginary pairs $\La=\pm\La_j\in i\mathbb{R}$

\item for pairs $\La_j^2,\La_j^{*\,2}$: quadruplets
 $\La=\pm\La_j,\pm\La_j^*\in \mathbb{C}\setminus\{\mathbb{R}\cup
 i\mathbb{R}\}$ .

\end{enumerate}
This leads to an accumulation of eigenvalues on both the real and the imaginary
axes as already pointed out in \cite{APSo}. The same phenomenon has been
observed numerically in a one-matrix model \cite{HOV} based on the proposal
\cite{Steph} (obtained from eq.\ (\ref{ZPQ}) by choosing $Q\sim I$).
This is in contrast to the real Ginibre model where eigenvalues
accumulate only on the real axis (see e.g. \cite{Sommers88}).

The joint probability distribution (jpdf) for the matrix $C$ is obtained from
eq.\ (\ref{ZAB}) by inserting a matrix delta function; using the cyclic
property of the trace we then have
\be
\fl
P(C)\sim  \exp[\eta_-\Tr\,C]
\int_{\mathbb{R}^{N(N+\nu)}}\!\!\!\!\!\!\!\!\!dA
\int_{\mathbb{R}^{N(N+\nu)}}\!\!\!\!\!\!\!\!\!dB
\exp\left[-\frac
{\eta_+}{2}\Tr(AA^T+BB^T)
\right]\delta(C-AB^T)\ .\
\label{PC}
\ee
As shown in Section \ref{jpdf}, our final result for the jpdf of
$\mcD$ in terms of squared variables $z_k = x_k + iy_k \equiv \La_k^2$
with $d^2z_k=dx_kdy_k$ is
\bea\fl
{\cal Z}&=& \int_{\mathbb{C}}
d^2z_1\ldots \int_{\mathbb{C}} d^2z_N P_N(z_1,\ldots,z_N)  \label{jpdfdef} \\
\fl\label{PM}
 & = & c_N\prod_{k=1}^{N} \int_{\mathbb{C}} d^2z_k\,w(z_k)\,
\prod_{i<j}^N(z_i-z_j)\sum_{n=0}^{[N/2]} \Big(
\prod_{l=1}^n
(-2i)\delta(x_{2l-1}-x_{2l})\delta(y_{2l-1}+y_{2l})\Theta(y_{2l-1})\nn\\
\fl&&\times
\Theta(x_1>x_3>...>x_{2n-1})
\Theta(x_{2n+1}>x_{2n+2}>...>x_N)\delta(y_{2n+1})...\delta(y_{N})
\Big).
\eea
The integration measure $d^2z_k$ extends over the complex plane for {\it each}
of the $z_k$. The normalisation constant $c_N$ will be given in eq.\
(\ref{CN}). In eq.\ (\ref{PM})
we sum over all distinct possibilities for $N$ eigenvalues to come in $n\geq0$
complex conjugate pairs, with the remaining $N-2n\geq0$ eigenvalues being
real. For $n=0$ in eq.\ (\ref{PM}) the product in the first line is simply
unity.

Specifically, the jpdf is only non-zero when the eigenvalues appear in the
following order: the $n$ complex eigenvalue pairs must be placed first,
ordered with
respect to decreasing real parts\footnote{
Unlike for real eigenvalues two complex eigenvalues that
are not complex conjugates can have the same real part, without the jpdf
vanishing. Although being of measure zero we can fix this ambiguity
by ordering with respect to decreasing absolute imaginary part.}
\bea
\fl
&&(\im \La_1^2>0),\ (\re \La_2^2=\re \La_1^2,\im \La_2^2=-\im
\La_1^2),\ (\re \La_3^2\leq\re \La_2^2,\im \La_3^2>0),\ldots,\nn\\
\fl
&&(\re \La_{2n-1}^2\leq\re \La_{2n-2}^2,\im\La_{2n-1}^2 > 0),\
(\re \La_{2n}^2=\re \La_{2n-1}^2,\im \La_{2n}^2=-\im\La_{2n-1}^2)
,
\label{comporder}
\eea
and the $N-2n\geq0$ real eigenvalues follow, and are also ordered with
respect to decreasing real parts:
\be
\La_{2n+1}^2>\La_{2n+2}^2>\ldots> \La_{N}^2\ .
\ee
The function $g(z)$ inside the weight function
\be
w(z)\equiv |z|^{\nu/2}\exp[\eta_-z]\,g(z)
\label{weight}
\ee
depends on whether $z\,(\equiv \La^2)$ is
real or complex:
\bea
\fl g(z) &\equiv&
2K_{\frac{\nu}{2}}(\eta_+|z|)\  ,\ \mbox{for}\ \ z\in \mathbb{R}\ ,
\label{fRdef} \\
\fl\left[g(z)\right]^2 &=&\left[g(z^*)\right]^2  \equiv
2\int_0^\infty \frac{dt}{t}\e^{-2\eta_+^2t(x^2-y^2)-\frac{1}{4t}}
K_{\frac{\nu}{2}}\left(2\eta_+^2t(x^2+y^2)\right)\erfc(2\eta_+\sqrt{t}|y|)
,\label{fCdef}
\\
\fl &&\mbox{for}\ z=x+iy\in \mathbb{C}.
\nn
\eea
Because the complex eigenvalues come in pairs we will always get {\it two}
factors $g(z)$ for each pair (note the square on the left hand side of
the definition eq.\ (\ref{fCdef})). The limit $y\to0$ of a single $g(z)$ in
eq.\ (\ref{fCdef}) is smooth, leading to eq.\ (\ref{fRdef}).

The essential idea in the derivation of the jpdf detailed in Section \ref{jpdf}
is to reduce the calculation of the jpdf for $\mcD$ with general $N$
down to $2\times2$ and $1\times1$ blocks, which can be handled in terms of
the $N=2$ and $N=1$ problems which we solve explicitly.

\subsection{Density correlation functions for finite $N$}

We follow the method of using a generating functional for all eigenvalue
density correlation functions introduced in
\cite{Sommers2007,Sommers2008}. Because we essentially follow
\cite{Sommers2007,Sommers2008} we can be brief here.
We enlarge the
definition of the partition function eq.\ (\ref{PM}) by introducing sources
$f(\La_k^2)$:\footnote{These will symmetrise the ordered eigenvalues when
  differentiating.}
\be
{\cal Z}[f]\equiv  \int_{\mathbb{C}}
d^2z_1\ldots \int_{\mathbb{C}} d^2z_N P_N(z_1,\ldots,z_N)
f(z_1)\ldots f(z_N)\ .
\label{Zg}
\ee
For pairwise distinct arguments $z_1 \neq z_2\neq\ldots \neq z_n$
the $n$-point density correlation functions are then generated in terms of
functional derivatives with respect to the sources,
leading to insertions of delta functions $\frac{\delta f(z)}{\delta
  f(z')}=\delta^2(z-z')$:
\be
R_n(z_1,\ldots,z_n)= \frac{\delta}{\delta f(z_1)}\ldots
\frac{\delta}{\delta f(z_n)}{\cal Z}[f]\Big|_{f\equiv1}\ .
\label{Rndef}
\ee
In doing so, each $n$-point function contains a sum of different
contributions, splitting $n$ into all possible combinations of
real eigenvalues and complex eigenvalue pairs.
Both the generating functional ${\cal Z}[f]$ and the $n$-point density $R_n$
can be written as Pfaffians \cite{Sinclair06,Sommers2007}:
\be
{\cal Z}[f]=c_N\ \Pf\left[ \int_{\mathbb{C}} d^2z_1 \int_{\mathbb{C}}
d^2z_2 \, f(z_1)f(z_2){\cal F}(z_1,z_2)
z_1^{i-1}z_2^{j-1}\right]_{1 \le i,j \le N},
\label{ZPf}
\ee
\bea
R_n(z_1,\ldots,z_n)&\equiv& \frac{N!}{(N-n)!}
 \int_{\mathbb{C}}
d^2z_{n+1}\ldots \int_{\mathbb{C}} d^2z_N P_N(z_1,\ldots,z_N) \nn\\
&=& \Pf\left[
\begin{array}{cc}
{\cal K}_N(z_i,z_j)& -G_N(z_i,z_j)\\
G_N(z_j,z_i) & -W_N(z_i,z_j)
\end{array}
\right]_{1 \le i,j \le n}.
\label{RnPf}
\eea
In the latter case, one has to compute the Pfaffian of the ordinary, $2n
\times 2n$ matrix composed of the matrix of quaternions inside the square
bracket.
We have restricted ourselves to even $N$ for simplicity. The case of odd
$N$ can be treated along the lines of \cite{Sommers2008}
(or \cite{ForresterMays} for an alternative formulation).
We have introduced the following functions of two complex variables
$z_j=x_j+iy_j$, $j=1,2$:
\bea
\fl{\cal F}(z_1,z_2)&=&  w(z_1)w(z_2)
\Big(2i\delta^2(z_1-z_2^*)\sgn(y_1)+\delta(y_1)\delta(y_2)
\sgn(x_2-x_1)\Big),
\label{Fdef}\\
\fl{\cal K}_N(z_1,z_2)&=&\frac{\eta_-}{8\pi(4\mu^2\eta_+)^{\nu+1}}
\nn\\\fl&&
\times\sum_{j=0}^{N-2}\left(\frac{\eta_-}{\eta_+}\right)^{2j}\!
\frac{(j+1)!}{(j+\nu)!}
\left\{
L_{j+1}^\nu\Big( \frac{z_2}{4\mu^2\eta_-}\Big)
L_{j}^\nu\Big( \frac{z_1}{4\mu^2\eta_-}\Big)
- (z_1\leftrightarrow z_2)
\right\}
\label{Kdef}\\
\fl&=&
\frac{\eta_-}{8\pi(4\mu^2\eta_+)^{\nu+1}}
\left(z_2\frac{\partial}{\partial z_2}- z_1\frac{\partial}{\partial z_1}
-\frac{z_2-z_1}{4\mu^2\eta_-}\right)
\nn\\\fl&&
\times
\sum_{j=0}^{N-2}\left(\frac{\eta_-}{\eta_+}\right)^{2j}\!
\frac{j!}{(j+\nu)!}
L_{j}^\nu\Big( \frac{z_1}{4\mu^2\eta_-}\Big)
L_{j}^\nu\Big( \frac{z_2}{4\mu^2\eta_-}\Big),
\label{Krel}\\
\fl G_N(z_1,z_2)&=&-\int_{\mathbb{C}} d^2z \, {\cal K}_N(z_1,z){\cal
  F}(z,z_2)\ ,
\label{Gdef}\\
\fl W_N(z_1,z_2)&=& -{\cal F}(z_1,z_2)+\int_{\mathbb{C}}  d^2z
\int_{\mathbb{C}} d^2z'{\cal F}(z_1,z)
{\cal K}_N(z,z'){\cal F}(z',z_2)\ .
\label{Wdef}
\eea
The kernel $\mcK_N(z_1,z_2)$ is the building block of all correlations,
and is actually defined in terms of $\mcF(z_1,z_2)$
(see Section \ref{dense} for further details). However, we will give a
different
proof of the precise form of eq.\ (\ref{Kdef}) which does not rely on direct
evaluation of the definition; rather, it will be calculated from the
expectation value of two characteristic
polynomials (see also \cite{APSo}).
Note that in eq.\ (\ref{Krel}) we have expressed the kernel of our real
two-matrix model ($\beta=1$) as a derivative of the kernel of the complex
two-matrix model \cite{Osborn} ($\beta=2$).
We found that a similar relation holds relating the kernel of
the $\beta=1$ real
Ginibre ensemble \cite{Forrester07,APSo} to the one for the $\beta=2$
complex Ginibre ensemble \cite{FKS98},
containing Hermite polynomials in the complex
plane.

There is an integration theorem in analogy to other matrix models with complex
eigenvalues \cite{Sommers2007,Mehta,TW}:
\bea
&& \int_{\mathbb{C}}  d^2z_n\Pf\left[
\begin{array}{cc}
{\cal K}_N(z_i,z_j)& -G_N(z_i,z_j)\\
G_N(z_j,z_i) & -W_N(z_i,z_j)
\end{array}
\right]_{1 \le i,j \le n}
\nn\\
&&= (N-n+1)
\Pf\left[\begin{array}{cc}
{\cal K}_N(z_i,z_j)& -G_N(z_i,z_j)\\
G_N(z_j,z_i) & -W_N(z_i,z_j)
\end{array}
\right]_{1 \le i,j \le n-1}.
\eea
This is the major advantage of working with an $n$-point correlation function
as defined in eq.\ (\ref{RnPf}) which contains all possible contributions
of real eigenvalues
and complex conjugate eigenvalue pairs.  Had we studied instead a
particular $n$-point function with a fixed and given number of
real and complex conjugate eigenvalues, then we would have found that
such a simple integration theorem
does not exist \cite{Akemann07}.

Let us spell out one example explicitly, the spectral density, which we will
give for even $N$ only. Details can be found in Section \ref{dense} including
figures, where we follow the method of
\cite{Sommers2007,Sommers2008}.
From eq.\ (\ref{RnPf}) we have
\be
R_1(z_1)= \int_{\mathbb{C}}  d^2z \, {\cal K}_N(z_1,z){\cal F}(z,z_1)
\equiv R_1^{\mathbb{C}}(z_1)+\delta(y_1)R_1^{\mathbb{R}}(x_1)\ .
\label{density}
\ee
Inserting the appropriate weight from eqs.\ (\ref{fRdef}) and (\ref{fCdef})
we obtain with $z=x+iy$
\bea
\fl
R_1^{\mathbb{C}}(z)&=& -2i|z|^{\nu}\e^{2\eta_- x}
\sgn(y){\cal K}_N(z,z^*)
\label{RcomplN}\\
\fl&&\times
2\int_0^\infty \frac{dt}{t}\exp\Big[-2\eta_+^2t(x^2-y^2)-\frac{1}{4t}\Big]
K_{\frac{\nu}{2}}\Big(2\eta_+^2t(x^2+y^2)\Big)
\erfc\left(2\eta_+\sqrt{t}|y|\right) ,
\nn\\
\fl
R_1^{\mathbb{R}}(x)&=&
\int_{-\infty}^\infty dx' \sgn(x-x')\,
|xx'|^{\frac{\nu}{2}}\e^{\eta_- (x+x')}
2K_{\frac{\nu}{2}}(\eta_+|x|)
2K_{\frac{\nu}{2}}(\eta_+|x'|)
\,{\cal K}_N(x,x').
\label{RrealN}
\eea
Eqs.\ (\ref{density}) -- (\ref{RrealN}) are valid for even $N$ only.
In the final step we can change from squared variables $z=x+iy=\La^2$ to Dirac
eigenvalues $\La$, using the simple transformations
\bea\label{densitytrans}
R_{1\, Dirac}^{\mathbb{C}}(z) & = & 4|z|^2 R_{1}^{\mathbb{C}}(z^2), \nn \\
R_{1\, Dirac}^{\mathbb{R}}(x) & = & 2|x| R_{1}^{\mathbb{R}}(x^2).
\eea
Note that the latter describes the density both of real Dirac eigenvalues, for
$R_{1\, Dirac}^{\mathbb{R}}(x)$ with $x\in\mathbb{R}$, and of purely imaginary
Dirac eigenvalues, for
$R_{1\, Dirac}^{\mathbb{R}}(x)$ with $x\in i\,\mathbb{R}$. Because
$R_{1}^{\mathbb{R}}(x^2)\neq R_{1}^{\mathbb{R}}(-x^2)$ this is not the same
function, see e.g. Fig. \ref{rhoRstrfigN} in Section \ref{dense}.

\subsection{The large-$N$ limit}\label{N}

In order to take the large-$N$ limit in principle
one first has to rescale all eigenvalues in
eq.\ (\ref{PM}) $\La_k\to \sqrt{N}\, \La_k$, which is equivalent to giving
all the matrix elements in eq.\ (\ref{ZPQ}) Gaussian weights $\exp[-(N/2)\Tr
P^T P]$ for $P$, and similarly for $Q$.
In this parametrisation the macroscopic spectral
density will have a compact support in the large-$N$ limit, given by a circle
for $\mu=1$, and an ellipse for $0<\mu<1$.
We will not discuss this macroscopic limit in detail but will focus on the
local correlations, i.e. the microscopic large-$N$ limit. Here one has to
distinguish between strong and weak  non-Hermiticity \cite{FKS},
each of which involves a second rescaling.

We first give the strong non-Hermiticity limit 
{\it after}
the first rescaling,
defined by keeping $\mu$ fixed and
only rescaling the eigenvalues according to
\be
\lim_{N\to\infty,\La\to0}N\La^2\equiv \la^2\ .
\label{strongdef}
\ee
This scaling actually cancels the first scaling.
Because of this the scaling limit is also true away from the origin.

We only give the microscopic kernel here, to be inserted into
eqs.\ (\ref{RcomplN}) and (\ref{RrealN}). For $\mu=1$ the kernel simplifies
to monic powers, see eq.\ (33) in \cite{APSo}. Its large-$N$ limit is easily
seen to lead to a modified Bessel function
\bea
\fl
\mcK^S_\nu(z_1,z_2)
&=&\frac{(z_1-z_2)}{64\pi 2^\nu}
\sum_{j=0}^\infty\frac{1}{j!(j+\nu)!}
\left(\frac{z_1 z_2}{4} \right)^{j}
=\frac{(z_1-z_2)}{64\pi}(z_1z_2)^{-\nu/2}I_\nu(\sqrt{z_1 z_2}).
\label{Kstr}
\eea
Because $\mu$ is not scaled here, in contrast to the weak limit below,
the insertion into eqs.\ (\ref{RcomplN}) and
(\ref{RrealN}) is relatively straightforward, apart from the phase for
negative real eigenvalues.
The case for general $\mu$ which can also be obtained by simply rescaling the
arguments in eq.\ (\ref{Kstr}) is discussed in Section \ref{dense} where we
also show plots.

The weak non-Hermiticity limit at the origin (again {\it after} the initial
rescaling above) is defined by scaling
both the squared
Dirac eigenvalues $\La^2$ and the chemical potential $\mu^2$ with
$2N$, corresponding to the volume in field theory:
\be
\lim_{N\to\infty,\mu\to0}2N\mu^2\equiv\al^2 \ ,\ \
\lim_{N\to\infty,\La\to0}(2N)^2\La^2\equiv \la^2\ .
\label{weakdef}
\ee
In this limit the macroscopic density is projected back onto the real
axis, with probability density given
by the semi-circle as for $\mu=0$ (when our model is Hermitian), whilst
microscopically the eigenvalues still extend into the complex plane.

The limiting microscopic kernel, as a function of squared variables
$z=\Lambda^2$, can be expressed in terms of our completely unscaled
finite-$N$ kernel as
\bea
\fl
\mcK^W(z_1, z_2) & \equiv & \lim_{N \rightarrow \infty} \left[
\frac{1}{(4N)^2}\left(\frac{z_1 z_2}{(4N)^2} \right)^{\nu/2}
  \mcK_{N}\left(\frac{z_1}{4N},\frac{z_2}{4N};\mu=\frac{\alpha}{\sqrt{2N}}
\right) \right] \nn \\
\fl & = & \frac{1}{256 \pi\al^2} \int_0^1 \,ds\, s^2\,\e^{-2\alpha^2 s^2}\,
\left\{\sqrt{z_1}\,J_{\nu+1}(s\sqrt{z_1})J_{\nu}(s\sqrt{z_2}) - (z_1
\leftrightarrow z_2) \right\}
\label{Kweak}
\eea
which is to be inserted into the definition of the density
eq. (\ref{density}).  
From eq.\ (\ref{RcomplN}) 
we obtain for the microscopic density of complex eigenvalues 
\bea
\fl
\rho_\nu^{\mathbb{C}\,W}(z)&=&
-2i\ \sgn(y)\ \exp\left[\frac{x}{4\al^2}\right]
{\cal K}^W(z,z^*) \nn\\
\fl&&\times
2\int_0^\infty \frac{dt}{t}
\exp\Big[-\frac{t(x^2-y^2)}{32\al^4}-\frac{1}{4t}\Big]
K_{\frac{\nu}{2}}\left(\frac{t(x^2+y^2)}{32\al^4}\right)
\erfc \Big(\frac{\sqrt{t}|y|}{4\al^2}\Big).
\label{RcomplS}
\eea
The real eigenvalue density is more subtle; we again refer to Section
\ref{dense} for a more detailed discussion of the weak limit, including
figures.

\sect{Calculation of the joint probability distribution function}\label{jpdf}

In this section we compute the joint probability density function (jpdf) as
stated in
eq.\ (\ref{PM}) for the squared non-zero eigenvalues of $\mcD$. For pedagogical
reasons we first compute the jpdf separately for $N=1$ and 2 in
Sections \ref{N1} and \ref{N2} respectively.
This is because we will need these results
when treating the general $N$ case in Section \ref{Ngen},
as these sub-blocks will appear in the computation of the general Jacobian.
Some technical details will be deferred to \ref{Jacobi}.
The cases with $N=1,2$ will make the parametrisation and residual
symmetries more transparent for later.

\subsection{The $N=1$ case}\label{N1}

In this simplest case our matrices $P$ and $Q$, or after changing variables
$A$ and $B$
in eq.\ (\ref{PQdef}), are of size $1\times(1+\nu)$ and are thus given by
vectors
$\textbf{a}$ and $\textbf{b}$, each of length $\nu+1$.
The eigenvalue equation (\ref{evrel}) for $\mcD$
becomes
\be
0=\La^\nu(\La^2-\textbf{a}\cdot\textbf{b})\ ,
\ee
and thus we only have a single non-zero (and real) eigenvalue $\La^2$ to
determine. Its (j)pdf is given by
\be
\fl
P(\Lambda^2) =
\frac{1}{(4\pi\mu)^{\nu+1}}
\int_{\mathbb{R}^{\nu+1}}\!\!\!\!\!\! d\textbf{a}
\int_{\mathbb{R}^{\nu+1}}\!\!\!\!\!\!  d\textbf{b}\,
\exp[ \eta_-\textbf{a}\cdot\textbf{b}]
\exp\left[-\frac{\eta_+}{2}\left(|\textbf{a}|^2+|\textbf{b}|^2\right) \right]
\delta(\Lambda^2-\textbf{a}\cdot\textbf{b}) \, .
\ee
We simplify this expression in two steps.
First, without loss of generality we may choose the
direction of $\textbf{b}$ as the first basis vector for $\textbf{a}$ in
Cartesian coordinates. This leads to a decoupling of the remaining components
$a_2,\ldots,a_{\nu+1}$, and the integral now only depends on $\textbf{b}$
through its modulus $b=|\textbf{b}|$.
Second, we choose polar coordinates
for the vector $\textbf{b}$, leading to the Jacobian $b^\nu$,
and, on symmetrically extending the integral over $b$ to
$-\infty$, we obtain
\bea
\label{N1ab}
\fl
P(\Lambda^2) & = &
\frac{1}{(4\pi\mu)^{\nu+1}}
\frac{2\pi^{(\nu+1)/2}}{\Gamma\left(\frac{\nu+1}{2}\right)}
\left(\frac{2\pi}{\eta_+}\right)^{\nu/2}\e^{ \eta_-\La^2}
\frac12\int_{-\infty}^\infty da_1 \int_{-\infty}^\infty db
\, \e^{-\frac12\eta_+(a_1^2+b^2)}
\delta(\Lambda^2 - a_1b) \, |b|^{\nu}
\nn \\
\fl
& = & \frac{1}{2^{3\nu/2+2}\mu^{\nu+1}\eta_+^{\nu/2}\sqrt{\pi}
\,\Gamma\left(\frac{\nu+1}{2}\right)}\ \e^{ \eta_-\La^2}\int_{-\infty}^\infty
db \, \e^{-\frac12\eta_+(b^2+\Lambda^4/b^2)} \, |b|^{\nu-1}.
\eea
The first new pre-factor comes from the surface area of the unit $\nu$-sphere
\be
S_\nu\equiv\frac{2\pi^{(\nu+1)/2}}{\Gamma\left(\frac{\nu+1}{2}\right)}
\ =\ \frac{VO(\nu+1)}{VO(\nu)}
\label{Sdef}
\ee
(with $VO(\nu)$ being the volume of the orthogonal group) through the
angular integration over $\textbf{b}$, the final pre-factor from the Gaussian
integrations over the decoupled components of $\textbf{a}$. It is important to
note 
that the first line of eq.\ (\ref{N1ab}) looks like a reduction to the $\nu=0$
case, 
apart from the extra factor $|b|^{\nu}$ from the Jacobian. We will use the same
strategy for $N=2$ in the following subsection. In the next step we change
variables $\e^t = b^2/|\Lambda^2|$ 
to arrive at
\bea
\fl
P(\Lambda^2)
& = &
\frac{1}{2^{3\nu/2+2}\mu^{\nu+1}\eta_+^{\nu/2}\sqrt{\pi}\,
\Gamma\left(\frac{\nu+1}{2}\right)} 
|\Lambda|^{\nu}  \e^{ \eta_-\La^2}\,2\,\int_{0}^{\infty} \cosh(\nu t/2)
\e^{-\eta_+|\Lambda^2| \cosh t} \, dt \nn \\
\fl
& = &
\frac{1}{2^{3\nu/2+2}\mu^{\nu+1}\eta_+^{\nu/2}\sqrt{\pi}\,
\Gamma\left(\frac{\nu+1}{2}\right)}
|\Lambda|^{\nu}  \e^{ \eta_-\La^2}
\,2\,K_{\frac{\nu}{2}}(\eta_+|\Lambda^2|)\ .
\label{N1jpdf}
\eea
Here we have used a particular representation eq.\ 9.6.24 in \cite{Abramowitz}
of the $K$-Bessel function. It directly gives $c_{N=1}$ times
the weight function in
eqs.\ (\ref{weight}) and (\ref{fRdef}) when changing variables $\La^2\to z$,
where 
\be
c_{N=1} =
\frac{1}{2\pi}\,\frac{1}{(2\pi)^{\nu/2}(2\mu)^{\nu+1}
\eta_+^{\nu/2}}\,\frac{S_{\nu}}{S_0}. 
\ee
This is consistent with eqs.\ (\ref{PM}), (\ref{weight}) and (\ref{fRdef}), and
ends our calculation for $N=1$.

As a remark, eq.\ (\ref{N1jpdf}) can be derived in various different ways,
including by Fourier transformation. It is known that if $a$ and $b$ are
independent random variables with normal distributions, then the product $c=ab$
has distribution function $P_0\sim K_0$. Consequently, the sum of $\nu+1$
(independent) such
variables $c_i$ has a distribution given by the convolution of $\nu+1$
functions $K_0$.
Fourier transformation $F$
turns this into an ordinary product, and so
we obtain
$P_{\nu} = F^{-1}\left\{\left( F(P_0)\right)^{\nu+1}\right\}\sim
K_{\frac{\nu}{2}}$, i.e.
on performing the integrals we again reach eq.\ (\ref{N1jpdf}).

\subsection{The $N=2$ case}\label{N2}

Our matrices $A$ and $B$ are now given by two row vectors $\textbf{a}_{j=1,2}$
and $\textbf{b}_{j=1,2}$ each of length $\nu+2$:
\be
\fl
A=\left(\begin{array}{c}
{\textbf{a}_{1}}\\
{\textbf{a}_{2}}
\end{array}\right)\ ,\ \
B=\left(\begin{array}{c}
{\textbf{b}_{1}}\\
{\textbf{b}_{2}}
\end{array}\right)\ ,\ \ C=AB^T=
\left(\begin{array}{cc}
\textbf{a}_{1}\cdot\textbf{b}_{1}& \textbf{a}_{1}\cdot\textbf{b}_{2}\\
\textbf{a}_{2}\cdot\textbf{b}_{1}& \textbf{a}_{2}\cdot\textbf{b}_{2}
\end{array}\right).
\label{ABN2def}
\ee
The eigenvalue equation (\ref{evrel})
\be
0=\La_\nu \det[\La^2I_2-AB^T]
\ee
has two solutions $\La_{1,2}^2$ which may (i) both be real or (ii) form a
complex conjugate pair. We will have to distinguish these two cases below.

For the  first step we reduce the calculation of $P(C)$ in eq.\ (\ref{PC})
to a matrix integral of $2\times 2$ matrices $A'$ and $B'$ times a Jacobian,
as we did for $N=1$ in the first line of eq.\ (\ref{N1ab}). The resulting
Jacobian here will be $\sim|\det B'|^\nu$.
Our aim is to rotate
both $\textbf{b}_{1}$ and $\textbf{b}_{2}$ into the $xy$-plane of the
coordinates for the $\textbf{a}_{j}$, as then
\be
\textbf{a}_i\cdot\textbf{b}_j =  \sum_{k=1}^{\nu+1}
a_{ik}b_{jk}
= a_{i1}b_{j1}' + a_{i2}b_{j2}'\ ,
\ee
and the remaining components of the $\textbf{a}_{j}$ decouple. Here
we use primed vectors and coordinates to denote the quantities after rotation.

For $\nu=1$ the corresponding Jacobian is obtained as follows. Rotating
the 3D vector into 2D by $\textbf{b}_{1}'=O\,\textbf{b}_{1}$ gives rise to a
factor $\sim |\textbf{b}_{1}|$. Alternatively it can be computed by comparing
the initial and final `volumes' (generalised surface areas, in fact), yielding
$\frac{S_2|\textbf{b}_1|^2}{S_1|\textbf{b}_1|} = 2|\textbf{b}_1|$,
where $S_{n}$ is given by eq.\ (\ref{Sdef}). The remaining rotation around
$\textbf{b}_1$ rotates $\textbf{b}_2$ into the $xy$-plane as well, giving
$\frac{S_1|\textbf{b}_2|\sin \theta}{S_0} = \pi |\textbf{b}_2|\sin \theta$,
in which $\theta \in [0,\pi]$ is the angle between $\textbf{b}_1$ and
$\textbf{b}_2$. The final Jacobian reads
\be
|\textbf{b}_1|\,|\textbf{b}_2|\sin \theta
= |\textbf{b}_1'|\,|\textbf{b}_2'|\sin \theta =
\left|\begin{array}{cc}
b_{11}' & b_{12}' \\
b_{21}' & b_{22}' \\
\end{array}\right|
=|\det B'| \ ,
\ee
where the last equality easily follows by parametrising
$\textbf{b}_1'$ and $\textbf{b}_2'$ in 2D, and the factor $\frac{S_2}{S_0}$
corresponds to the angular integration over $\textbf{b}$.

For $\nu>1$ we can thus successively repeat these steps by projecting onto one
dimension lower until we reach the $xy$-plane. The volume factors will
telescope out and we arrive at
\be
\frac{S_{\nu}S_{\nu+1}}{S_0 S_1}(|\textbf{b}_1'||\textbf{b}_2'|\sin
\theta)^\nu=\frac{S_{\nu}S_{\nu+1}}{S_0 S_1}|\det B'|^\nu.
\ee
We thus have reduced eq.\ (\ref{PC}) for $N=2$ from $2\times(2+\nu)$ down to
$2\times 2$ matrices:
\bea\label{PC22}
\fl
P(C) &=& \frac{c_{N=2}\,\eta_+}{2\pi^3}
\,\e^{\eta_-\Tr\,C}\\
\fl&&
\times\int_{\mathbb{R}^4}\!\! dA'\int_{\mathbb{R}^4}\!\! dB'\,
\exp\left[
-\frac12\eta_+\Tr(A'A'^T+B'B'^T)\right] \delta\left( C - A'B'^{T}
\right)\,|\det B'|^{\nu}\ ,
\nn
\eea
where $c_{N=2}$ is defined in anticipation of the final result as
\be
c_{N=2} =
\frac{1}{8\pi}\,\frac{1}{(2\pi)^{\nu}(2\mu)^{4+2\nu}
\eta_+^{\nu+1}}\,\frac{S_{\nu}S_{\nu+1}}{S_0S_1}.
\ee
The Gaussian integrals over the decoupled components of the two
vectors $a_{jk}$ for $k>2$ have been evaluated, using that
$\Tr AA^T=|\textbf{a}_1|^2+|\textbf{a}_2|^2$. The $2\times 2$
matrix $A'$ can now be
integrated out, by formally changing variables $A'\to F=A'B'^T$ with Jacobian
$|\det B'|^{-2}$:
\bea
\fl
P(C) = \frac{c_{N=2}\,\eta_+}{2\pi^3} \,
\e^{\eta_-\Tr\,C}\!
\int_{\mathbb{R}^4}\!\! dB'\,
\exp\left[
-\frac12\eta_+\Tr\left(CC^T(B'B'^T)^{-1}+B'B'^T\right)\right] |\det
B'|^{\nu-2}. \nn\\
\fl\label{PCN2red}
\eea
Note the similarity with eq.\ (\ref{N1ab}).

In a second step we perform the integral $\int dB'$. Because $CC^T$ is a
symmetric, positive definite matrix we can diagonalise it with an orthogonal
transformation $O$
\be
O^T (CC^T)O=\left(\begin{array}{cc}
\la_1 & 0 \\
0 & \la_2
\end{array}\right),\ \ \la_{1,2}\geq0\ .
\label{Csymdiag}
\ee
Using the invariance of $dB'$ we can change variables $B'\to OB'$ with
$(B'B'^T)^{-1}\to O(B'B'^T)^{-1} O^T$. We thus
replace $CC^T$ by its diagonalised form
eq.\ (\ref{Csymdiag}) in the exponent in eq.\ (\ref{PCN2red}):
\be
\fl\Tr\left(B'B'^T+CC^T(B'B'^T)^{-1}\right)=a^2 + b^2 + c^2 + d^2
+h^{-2} \big( (c^2+d^2)\lambda_1 + (a^2+b^2)\lambda_2 \big),
\ee
where we have explicitly parametrised
\be
B' \equiv \left( \begin{array}{cc} a & b \\ c & d \end{array}\right)
,\ h\equiv\det B'=ad-bc \ .
\ee
We now introduce $h$ as an independent variable in eq.\ (\ref{PCN2red})
by inserting a delta-function
constraint in its integral representation:
\bea
\fl
P(C)&=& \frac{c_{N=2}\,\eta_+}{4\pi^4}\,
\e^{\eta_-\Tr\,C}\, \int_{-\infty}^\infty dh \, |h|^{\nu-2}
\int_{\mathbb{R}^4}\!\!dB' \,
\int_{-\infty}^\infty d\omega\,\e^{-i\omega (h-(ad-bc))} \\
\fl&& \times \exp\left[ -\frac{\eta_+}{2}[a^2+b^2+c^2+d^2
  + h^{-2}((c^2+d^2)\lambda_1+(a^2+b^2)\lambda_2)] \right]
\nn\\
\fl & = & \frac{c_{N=2}\,\eta_+}{4\pi^4}\, \e^{\eta_-\Tr\,C}\,
2\int_{0}^{\infty} dh
\, h^{\nu-2} \,
\int_{-\infty}^{\infty} d\omega \, \e^{-i\omega h}
\frac{(2\pi)^2}{\omega^2+\eta_+^2(1+\la_1/h^2)(1+\la_2/h^2)},\nn
\eea
where we performed the Gaussian integrals successively in pairs $a,d$ and
$b,c$, and switched to positive $h$.
The denominator in the third line
can be rewritten at the cost of an additional integral,
$\frac{1}{a} = \int_{0}^{\infty}dt\,\e^{-at}$ for $a>0$, and after changing
variables $\omega\to\tau=\omega h$ we have
\bea
\fl
P(C)&=& \frac{2c_{N=2}\,\eta_+}{\pi^2}\,
\e^{\eta_-\,\Tr\,C}
\int_{0}^{\infty} dh \, h^{\nu-1} \, \int_{-\infty}^{\infty} d\tau \,
\e^{-i\tau}
\int_{0}^{\infty}dt\,
\e^{-(\tau^2+\eta_+^2(h^2+(\lambda_1+\lambda_2)+\lambda_1\lambda_2/h^2))t}\nn\\
\fl&=& \frac{2c_{N=2}\,\eta_+}{\pi^{3/2}}  \, (\lambda_1 \lambda_2)^{\nu/4}
\,\e^{\eta_-\Tr\,C}
\int_{0}^{\infty} \frac{dt}{\sqrt{t}} \, \exp\left[
-\eta_+^2(\lambda_1+\lambda_2)t - \frac{1}{4t} \right] \,
K_{\frac\nu2}(2\eta_+^2 \sqrt{\lambda_1 \lambda_2} t).\nn\\
\fl
\label{PCsymN2}
\eea
We performed the Gaussian integration over $\tau$ first, and then employed the
following integral from \cite{Grad} eq.\ 8.432.6
\be
K_{\nu}(z) = \frac{1}{2}\,\left( \frac{z}{2} \right)^{\nu}\, \int_{0}^{\infty}
dt\,\frac{\e^{-t-z^2/4t}}{t^{\nu+1}}
\ee
to do the $h$-integration after another change of variables,
arriving at our result eq.\ (\ref{PCsymN2}).

As the final step we need to express $P(C)$ in terms of the eigenvalues of $C$
(i.e. $\La_1^2$ and $\La_2^2$), rather than those of $CC^T$. This will also
involve a Jacobian to be computed later. In eq.\ (\ref{PCsymN2})
we need
\be
\Tr\,C=\La_1^2+\La_2^2\ ,\qquad \la_1\la_2=\det[CC^T]=\La_1^4\La_2^4\ ,
\label{laprod}
\ee
which are trivial. Only the combination
$\lambda_1+\lambda_2= \Tr (CC^T)$
requires more calculation. In general, we can orthogonally transform any real
matrix $C$ to the form
\be
C=\left( \begin{array}{cc} \sin\theta & -\cos\theta \\ \cos\theta & \sin\theta
\end{array} \right)
\left( \begin{array}{cc} \epsilon_1 & s \\ -s & \epsilon_2 \end{array} \right)
\left( \begin{array}{cc} \sin\theta & \cos\theta \\ -\cos\theta & \sin\theta
\end{array} \right)
\label{Cpara}
\ee
where $\epsilon_1$, $\epsilon_2$, $s$ and the rotation parameter $\theta$ are
all real. The matrix parameters $\{\epsilon_1,\epsilon_2,s\}$ and
eigenvalues  $\{\Lambda_1^2, \Lambda_2^2\}$ follow from
\be
0 = \left| \begin{array}{cc} \epsilon_1-\Lambda^2 & s \\ -s &
  \epsilon_2-\Lambda^2 \end{array} \right| = \
\Lambda^4 - (\epsilon_1+\epsilon_2)\Lambda^2 + (\epsilon_1\epsilon_2+s^2)\ ,
\ee
with solution
\be
\Lambda_{1,2}^2 = \frac{\epsilon_1+\epsilon_2}{2} \pm
\sqrt{\frac{(\epsilon_1-\epsilon_2)^2}{4}-s^2}.
\ee
This can be inverted for $\epsilon_{1,2}$ for later use to
\be\label{epsdef}
\epsilon_{1,2} = \frac{\Lambda_1^2+\Lambda_2^2}{2} \pm
\sqrt{\frac{(\Lambda_1^2-\Lambda_2^2)^2}{4}+s^2}\ .
\ee
This immediately leads to
\be
\lambda_1+\lambda_2= \Tr (CC^T)= \epsilon_1^2 + \epsilon_2^2 + 2s^2=
\Lambda_1^4+ \Lambda_2^4 + 4s^2.
\label{lasum}
\ee
On inserting eqs.\ (\ref{laprod}) and (\ref{lasum}) into eq.\ (\ref{PCsymN2})
we 
can express $P(C)$ in terms of $\La_{1,2}^2$. However, we have changed
variables twice to arrive here, first from the matrix elements $c_{ij}$ of $C$
to $\{\epsilon_1,\epsilon_2,s,\theta\}$, and second from
$\{\epsilon_1,\epsilon_2\}$ to $\{\La_1^2,\La_2^2\}$.
The corresponding Jacobians we must multiply by are given by
\bea
{\cal J}_1&=&
\left|\frac{\partial\{c_{11},c_{12},c_{21},c_{22}\}}
{\partial\{\epsilon_1,\epsilon_2,s,\theta\}}\right|
=2\,|\epsilon_1-\epsilon_2|\,\cos^2(2\theta)
\ ,\nn\\
{\cal J}_2 &=& \left|\frac{\partial\{\epsilon_1,\epsilon_2\}}
{\partial\{\Lambda_1^2,\Lambda_2^2\}}\right|=
\left| \frac{\Lambda_1^2-\Lambda_2^2}{\epsilon_1-\epsilon_2} \right| \ .
\eea

\subsubsection{Distinction of real and complex eigenvalues}\ \\[-2ex]

In order to give the jpdf from eq.\ (\ref{PCsymN2}) for variables $\La_{1,2}^2$
alone we have to integrate over the remaining real variables $s$ and $\theta$.
Here we have to distinguish between the case of two real eigenvalues, and that
of a complex conjugate pair. 
Starting from a real matrix $C$ in eq.\ (\ref{Cpara}) all new variables, in
particular $\epsilon_{1,2}$, are real. In view of eq.\ (\ref{epsdef}) there are
two possibilities if the radicand is to remain positive:
\begin{itemize}
\item[(i)] $\Lambda_1^2,\Lambda_2^2$ are both real $\Rightarrow
  \frac14(\Lambda_1^2-\Lambda_2^2)^2+s^2 \ge 0$
which is always satisfied, or

\item[(ii)] $\Lambda_1^2,\Lambda_2^2$ are complex conjugates:
$\Lambda_1^2-\Lambda_2^2\in i\mathbb{R}$
$\Rightarrow s^2 \ge - \, \frac14(\Lambda_1^2-\Lambda_2^2)^2 \ge 0$.

\end{itemize}
We thus obtain for the jpdf
\bea
\fl
d\La_1^2 d\La_2^2 P(\La_1^2,\La_2^2) &=&  \frac{2c_{N=2}\,\eta_+}{\pi^{3/2}}
\,
d\La_1^2 d\La_2^2\,
 \e^{\eta_-(\Lambda_1^2+\Lambda_2^2)}
 |\Lambda_1^2
\Lambda_2^2|^{\nu/2}
\int_0^{2\pi}d\theta
\left|\Lambda_1^2-\Lambda_2^2 \right| \cos^2(2\theta) \nn\\
\fl&\times&\int_{s_{min}}^\infty ds \,
2 \int_{0}^{\infty} \frac{dt}{\sqrt{t}}
\exp\left[-\eta_+^2(\Lambda_1^4+\Lambda_2^4+4s^2)t - \frac{1}{4t} \right]
K_{\frac\nu2}(2\eta_+^2 |\Lambda_1^2 \Lambda_2^2| t),\nn\\
\fl
\label{preP}
\eea
where $s_{min}^2=\max\{0,-(\Lambda_1^2-\Lambda_2^2)^2/4  \}$. The
$\theta$-integral is trivial.

For two real eigenvalues having $s_{min}=0$ the integral over $s$ can be
performed, leading to the following simplification for the remaining
$t$-integral:
\be
\fl
2\,\int_{0}^{\infty} \frac{dt}{t}\,
\e^{ -\eta_+^2(\Lambda_1^4+\Lambda_2^4)t - \frac{1}{4t}}
K_{\frac\nu2}(2\eta_+^2 |\Lambda_1^2 \Lambda_2^2| t)
=
2K_{\frac\nu2}(\eta_+|\Lambda_1^2|)\,2K_{\frac\nu2}(\eta_+|\Lambda_2^2|)\ .
\label{Kid}
\ee
Here we have used \cite{Grad} eq.\ 6.653.2 after changing variables
$t\to u=\frac{1}{2t}$. When ordering the two eigenvalues as $\La_1^2>\La_2^2$
the jpdf eq.\ (\ref{preP}) can thus be written as
\be
\fl
P(\La_1^2,\La_2^2) = c_{N=2} \, (\Lambda_1^2-\Lambda_2^2)\prod_{j=1,2}
|\Lambda_j^2|^{\nu/2}\e^{\eta_-\Lambda_j^2}\ 2
K_{\frac\nu2}(\eta_+|\Lambda_j^2|), \ \ \La_{1,2}^2\in\mathbb{R}\ ,
\label{PN2real}
\ee
as was claimed in eq.\ (\ref{PM}) in conjunction with eq.\ (\ref{fRdef}) for
$N=2$.

For two complex conjugate eigenvalues the $s$-integral leads to the
complementary error function, without further simplification. Here we order
$0<\im\La_1^2=-\im \La_2^2$ as in eq.\ (\ref{comporder}) to obtain the
following real, positive distribution
\bea
\fl
P(\La_1^2,\La_2^2) \, d\La_1^2d\La_2^2 &=& c_{N=2} \, d\La_1^2d\La_2^2 \,
(\Lambda_1^2-\Lambda_2^2)
|\Lambda_1^2|^{\nu/2}|\Lambda_2^2|^{\nu/2}\e^{\eta_-(\Lambda_1^2+\La_2^2)}
\label{PN2comp}
\\
&&\fl\quad\quad\times
2\int_{0}^{\infty} \frac{dt}{t} \,
\e^{ -\eta_+^2(\Lambda_1^4+\Lambda_2^4)t - \frac{1}{4t}}
K_{\frac\nu2}(2\eta_+^2 \La_1^2\La_2^2 t) \,
\erfc(\eta_+\sqrt{t}\,|\Lambda_1^2-\Lambda_2^2| ) ,
\ \ \La_{1,2}^2\in\mathbb{C}\ .
\nn
\eea
Because the integral depends only on the modulus $|\im\La_1^2|=|\im \La_2^2|$
we can define its square root to be the weight of each eigenvalue $\La_j^2$,
as in eq.\ (\ref{fCdef}). The limit  $|\im\La_1^2|\to0$ smoothly reduces the
integral in
eq.\ (\ref{PN2comp}) to eq.\ (\ref{Kid}), using that $\erfc(0)=1$.

Combining the real and complex cases, and still assuming the eigenvalue
ordering, we can write the jpdf most generally as follows
\bea\fl
d^2z_1d^2z_2P(z_1,z_2) &=&c_{N=2}\,d^2z_1d^2z_2(z_1-z_2) w(z_1)
w(z_2) \label{PN2gen}\\
\fl&&\times\Big( \delta(y_1)\delta(y_2)\Theta(x_1-x_2)
- 2i\delta^2(z_1-z_2^{*}) \Theta(y_1)\Big)
\nn
\eea
where we have switched variables $\La^2=z=x+iy$, including for the
differentials, $d\La^2d\La^{*\,2}=(dx+idy)(dx-idy)=-2idxdy = -2id^2z$.
This is just the jpdf in eq.\ (\ref{PM}) including the weight
eq.\ (\ref{weight}), and this completes our computation for the $N=2$ case.

\subsection{General structure for arbitrary $N$}\label{Ngen}

In this subsection we will compute the jpdf for any $N$ -- both even and odd --
given in terms of the eigenvalues
$\La_j^2$ of the Wishart matrix $C=AB^T$. We start from eq.\ (\ref{PC})
which we repeat here for convenience,
\be
\fl
P(C)\sim  \exp[\eta_-\Tr\,C]\int_{\mathbb{R}^{N(N+\nu)}}\!\!\!\!\!\!\!\!dA
\int_{\mathbb{R}^{N(N+\nu)}}\!\!\!\!\!\!\!\!dB
\ \exp\left[-\frac
{\eta_+}{2}\Tr(AA^T+BB^T)
\right]\delta(C-AB^T)\ .\
\label{PC2}
\ee
We will eventually use the results from the previous two subsections as
building blocks.

Instead of using a generalised Schur (or $QZ$) decomposition involving
unitary matrices to bring $A$ and $B$ to upper triangular form, we will
restrict ourselves here to orthogonal transformations.
The best we can achieve in this way is a so-called almost (or quasi) upper
triangular (AUT) form for one of the matrices, and an upper triangular form for
the second (which is also AUT). An AUT matrix is composed of a block diagonal
matrix, having non-vanishing $2\times2$
blocks along the diagonal for even $N$, and an additional $1\times1$ block at
the end if $N$ is odd. The remaining non-zero elements of an AUT matrix
all lie above the block diagonal.

The precise transformation that we make is:
\be\label{ABtrafo}
A = O_A (\Delta_A + \Lambda_A) O_B^{T},
\qquad B^T = O_B (\Delta_B +\Lambda_B) O_A^{T}.
\ee
Here $\Lambda_A$ and $\Lambda_B$ are block diagonal,
and $\Delta_A$ and $\Delta_B$ are zero except in elements strictly above the
block diagonal. $\Delta_A+\Lambda_A$ and $\Delta_B+\Lambda_B$ are hence AUT.
Note that $O_A$ is of size $N \times N$, and $O_B$ is of size $(N+\nu) \times
(N+\nu)$. $\Delta_A$ and $\Lambda_A$ are each the same size as $A$ itself
(i.e. rectangular), and similarly $\Delta_B$ and $\Lambda_B$ are the same size
as $B^T$.

To make the transformation eq.\ (\ref{ABtrafo}) unique, having the same number
of degrees of freedom (dof) on the left- and right-hand sides,
we restrict $O_A$ and $O_B$ as follows:
\bea
\fl
{O}_A & \in  O(N)/O(2)^{N/2} \ ,\
&{O}_B   \in  O(N+\nu)/O(2)^{N/2} \, O(\nu)
\ \ \ \ \ \ \, \qquad \mbox{for even $N$}\ ,\label{coset}  \nn \\
\fl
{O}_A & \in  O(N)/O(2)^{(N-1)/2}\ ,\
&{O}_B   \in  O(N+\nu)/O(2)^{(N-1)/2}\,O(\nu)
\ \ \qquad \mbox{for odd $N$}.
\label{cosetodd}
\eea
The residual symmetries leading to these cosets are rotations within each
block on the diagonal, and, loosely speaking, the extra factor $O(\nu)$ can be
thought of as originating 
from the reduction of $B$ from rectangular to square form, as in the $N=1$ and
2 cases earlier. The precise counting of dof is given in Table \ref{t1},
matching the sum of 
dof of $A$ and $B$ for all $N$.

\begin{table}
\begin{center}\begin{tabular}{c|l||c|l|l}
Matrix & Degrees of freedom & Matrix & \multicolumn{2}{c}{Degrees of freedom}
       \\
       &                    &        & Even $N$ & Odd $N$ \\
\hline
 & & & \\
$A$ & $N(N+\nu)$ & $\Lambda_A$ & $2N$ & $2N-1$ \\
 & & & \\
$B$ & $N(N+\nu)$ & $\Lambda_B$ & $2N$ & $2N-1$\\
 & & & \\
 & & $\Delta_A$ & $\frac{N^2}{2}-N+\nu N$ & $\frac{N^2+1}{2} - N+ \nu N$ \\
 & & & \\
 & & $\Delta_B$ & $\frac{N^2}{2}-N$ & $\frac{N^2+1}{2} - N $ \\
 & & & \\
 & & ${O}_A$ & $\frac{N^2}{2}-N$ & $\frac{N^2+1}{2} - N$ \\
 & & & \\
 & & ${O}_B$ & $\frac{N^2}{2}-N+\nu N$ & $\frac{N^2+1}{2} - N + \nu N$ \\
 & & & \\
\end{tabular}\end{center}
\caption{\label{t1}
Counting dof before and after change of variables.}
\end{table}

Under this orthogonal transformation, the integrand in eq.\ (\ref{PC})
changes as follows:
\be
\fl
\exp\left[-\frac{\eta_+}{2}\Tr(AA^T + BB^T)\right] =
\exp\left[-\frac{\eta_+}{2}\Tr(\Lambda_A
\Lambda_A^T + \Lambda_B \Lambda_B^T + \Delta_A\Delta_A^T + \Delta_B
\Delta_B^T)\right].
\ee
The pre-factor
$\exp\left[\eta_-\Tr C\right]=\exp\left[\eta_- \sum_{i=1}^N \Lambda_i^2\right]$
remains unchanged, with the relation between the eigenvalues $\La_i^2$ and the
new variables yet to be determined.

The transformation eq.\ (\ref{ABtrafo}) leads to the following differentials
\bea
\fl
(dA)_{ij} & = & \Big( O_A \big[ O_A^{T}dO_A(\Delta_A+\Lambda_A) -
  (\Delta_A+\Lambda_A)O_B^{T}dO_B +d\Delta_A + d\Lambda_A \big] O_B^{T}
\Big)_{ij}, \nn
\\
\fl
(dB^T)_{ij} & = & \Big( O_B \big[ O_B^{T}dO_B(\Delta_B+\Lambda_B) -
  (\Delta_B+\Lambda_B)O_A^{T}dO_A +d\Delta_B + d\Lambda_B \big] O_A^{T}
\Big)_{ij}.
\label{dAdB}
\eea
Here we have used the fact that for orthogonal transformations (with $O^TO=I$)
the differential $O^TdO$ is anti-symmetric. This remains true for our special
choice of cosets.
When considering the invariant line element $\Tr(dA\, dA^T+dB\,dB^T)$ the
rotations outside the square brackets in eq.\ (\ref{dAdB}) can be dropped.

We will now compute the Jacobian for the change of variables from
$\{dA,dB^T\}$ to
 $\{d\Lambda_A,d\Lambda_B,d\Delta_A,d\Delta_B,O_A^TdO_A,O_B^TdO_B\}$.
Here we use the differentials for convenience as for the orthogonal matrices
only these constitute independent variables (see also \cite{Mehta} for
a similar discussion for the real Ginibre ensemble).
In particular, counting dof
$O_A^TdO_A$ is an anti-symmetric matrix with zeros not just on the diagonal but
on the block diagonal. Similarly $O_B^TdO_B$ is an anti-symmetric matrix with
zeros on the block diagonal for the first $N$ elements, and a zero-block of
size $\nu\times\nu$ on the remaining part of the diagonal. 

In the differential eq.\ (\ref{dAdB}) the variables
$\{d\Lambda_A,d\Lambda_B,d\Delta_A,d\Delta_B\}$ are already diagonal, with
\be
\frac{\partial (dA)_{ij}}{\partial (d\Delta_A)_{pq}} = \delta_{ip} \delta_{jq}
\ ,\ \
\frac{\partial (dA)_{ij}}{\partial (d\Lambda_A)_{pq}} =  \delta_{ip}
\delta_{jq} \ ,
\ee
and similarly for $dB^T$. This contributes a unity matrix block to the
Jacobi matrix ${\cal J}$, when considering the corresponding elements of
$\{dA,dB^T\}$ on and above the block diagonal.

The non-trivial contribution from the Jacobian therefore originates from
differentiating the remaining elements of $\{dA,dB^T\}$ below
the block diagonal with respect to the independent variables of
$\{O_A^TdO_A,O_B^TdO_B\}$, where we also choose the lower  block diagonal
elements for convenience.
When appropriately ordering the elements of the
Jacobi matrix (see \ref{Jacobi} and also a similar
discussion in \cite{Mehta}) the contributions
proportional to $\Delta_A$ and $\Delta_B$ in eq.\ (\ref{dAdB}) will drop out,
being part of a lower triangular sub-matrix in ${\cal J}$.

For the sake of argument we restrict ourselves in this section
to the case of $\La_A$ and $\La_B$ being diagonal.
The more general (and typical) case in which $\Lambda_A$ and $\Lambda_B$
contain $2\times2$ blocks is treated in \ref{Jacobi}.

Arranging the remaining matrix elements below the block diagonal of $dA$ and
of the square part of $dB^T$
into pairs this leads to a $2\times2$ block diagonal Jacobi sub-matrix with
elements
\bea
\fl
\left|
\det\left(\begin{array}{cc}
	\frac{\partial (dA)_{ij}}{\partial(O_A^{T}d O_A)_{ij}}
& \frac{\partial (dB)_{ij}}{\partial (O_A^{T}d O_A)_{ij}} \\
	\frac{\partial (dA)_{ij}}{\partial (O_B^{T}d O_B)_{ij}}
& \frac{\partial (dB)_{ij}}{\partial (O_B^{T}d O_B)_{ij}}
\end{array}\right)
\right|
& = &
\left|
\det\left(\begin{array}{cc}
	 (\Lambda_A)_{jj} & -(\Lambda_B)_{ii} \\
	-(\Lambda_A)_{ii} & (\Lambda_B)_{jj}
\end{array}\right)
\right|
\ =\ |\Lambda_j^2 - \Lambda_i^2|\ ,
\label{Ladiag}
\eea
where we used that $(\Lambda_A)_{jj}(\Lambda_B)_{jj} = \Lambda_j^2$ (no
summations), and the $\Lambda_j^2$ are the eigenvalues of the matrix $C \equiv
AB^T$. The remaining $\nu N$
matrix elements below the block diagonal of $dB^T$ give a diagonal sub-matrix
with $\nu$ elements $(\Lambda_B)_{jj}$ each. The resulting contribution to
the Jacobian is
\be
{\cal J}= \prod_{1\leq i<j\leq N}'
|\Lambda_i^2-\Lambda_j^2|\
\prod_{j=1}^{N} |(\Lambda_B)_{jj}|^{\nu} \ .
\ee
The prime on the product symbol denotes that only those factors with indices
$(i,j)$
strictly below the block diagonal are to be included.
Finally, we observe that the second product can be written as
\be\label{2nd_prod_result}
\prod_{j=1}^{N} |(\Lambda_B)_{jj}|^{\nu} = \left\{
\begin{array}{ll}
			\prod_{j=1}^{N/2} |\det{B_j}|^{\nu}
                         &\ \ \textup{for even }N,  \\ \\
			\prod_{j=1}^{(N-1)/2}
			|\det{B_j}|^{\nu}|(\Lambda_B)_{NN}|^{\nu}
			&\ \ \textup{for odd }N,
\end{array}
\right.  
\ee
in which $B_j$ is the $j$-th $2 \times 2$ block along the diagonal of the
matrix $\Lambda_B$.
The same statement is true in the more general case when $\Lambda_A$ (and
$\Lambda_B$) are not diagonal, as shown in \ref{Jacobi}.

Writing everything together, we have for the total measure
\bea
\fl&& dA \, dB\ \e^{\eta_-\Tr\,AB^T-\frac{\eta_+}{2}\Tr(AA^T+BB^T)}\nn
\label{diff}\\
\fl&\sim& d\Lambda_A\,d\Lambda_B\,d\Delta_A\,d\Delta_B\,O_A^{-1}dO_A\,
O_B^{-1}dO_B\,
\prod_{1\leq i<j\leq N}^{\prime} |\Lambda_i^2-\Lambda_j^2|
\prod_{i=1}^{N} \e^{\eta_- \Lambda_i^2} \\
\fl&&\times
\e^{-\frac{\eta_+}{2}\Tr(\Lambda_A^T \Lambda_A + \Lambda_B^T \Lambda_B +
  \Delta_A^T \Delta_A + \Delta_B^T \Delta_B)}
\left\{
\begin{array}{ll}
			\prod_{j=1}^{N/2} |\det{B_j}|^{\nu} &\ \textup{for
			even }N   \\
			\prod_{j=1}^{(N-1)/2}
			|\det{B_j}|^{\nu}|\Lambda_B|_{NN}^{\nu}&\
			\textup{for odd }N.
\end{array}
\right.  
\nn
\eea
All constant factors are omitted here; we give the overall normalisation
constant later. 
We reiterate that $B_j$ here is the $j$-th $2 \times 2$ block on the diagonal
of 
$\Lambda_B$. The integration over the orthogonal dof as well as over the upper
block 
triangular matrices $\Delta_{A,B}$ can now be performed as they decouple. The
relevant dof for the right-hand side of eq.\ (\ref{diff}) can thus be written
in terms of the $\La_i^2$ and $2\times2$ blocks of the matrices
$\Lambda_{A,B}$
\bea
\fl
\prod_{i<j}^{\prime} |\Lambda_i^2-\Lambda_j^2|
\left\{
\begin{array}{ll}
			\prod_{i=1}^{N/2} \left\{
			dA_i\,dB_i\,
\e^{\eta_-(\Lambda_{2i-1}^2+\Lambda_{2i}^2)}\e^{-\frac{\eta_+}{2}
\Tr(A_i^TA_i+B_i^TB_i)}\,|\det B_i|^{\nu}  \right\} &\textup{for even }N,   
\\ \\
			\prod_{i=1}^{(N-1)/2} \big\{ \textup{ ditto }
			\big\}_{i}\ da\,db\ \e^{\eta_-
			\Lambda_N^2}\e^{-\frac{\eta_+}{2}(a^2+b^2)}|b|^{\nu}
			&\textup{for odd }N.
\end{array}
\right.  
\nn
\eea
\be
\label{prejpdf}
\ee
We have thus reduced the problem of computing the jpdf to a simpler problem
involving
 only $2 \times 2$ and $1 \times 1$ blocks, which can be handled just as the
 $N=2$ and $N=1$ cases that we treated in the previous two
 subsections. For this, it is necessary to order the eigenvalues as described
 in eq.\ (\ref{comporder}). Then, following eq.\ (\ref{PN2gen}), each $2
 \times 2$ block will make 
 the following contribution in variables $\La_i^2=z_i$
(with $\Lambda_{2i-1}^2$ and $\Lambda_{2i}^2$ either both real or complex
 conjugates, and 
ordered as described at the end of Section \ref{N2})
\bea
&&d^2z_{2i-1}\,d^2z_{2i}(z_{2i-1} - z_{2i})w(z_{2i-1})w(z_{2i})\nn\\
&&\times
\Big( \delta(y_{2i-1})\delta(y_{2i})\Theta(x_{2i-1}-x_{2i})
- 2i\delta^2(z_{2i-1}-z_{2i}^{*})\Theta(y_{2i-1}) \Big)\,.
\eea
In a similar way, we have for the $1 \times 1$ block when $N$ is odd
\be
d^2z_N\,|z_N|^{\nu/2}\,\e^{\eta_- z_N}\,g(z_N)\,\delta(y_N)
=d^2z_Nw(z_N)\,\delta(y_N)\,,
\ee
where $z_N$ will be real. Note that collecting all these quantities in eq.\
(\ref{prejpdf}) for each $i=1$ to $[N/2]$ the factors
$\Lambda_{2i-1}^2-\Lambda_{2i}^2$  will combine with the
$\prod_{i<j}^{'}(\Lambda_{i}^2-\Lambda_{j}^2)$ to make a true Vandermonde
determinant $\prod_{i<j}(\Lambda_{i}^2-\Lambda_{j}^2)$,
and we were able to drop the modulus sign because of the chosen ordering.
The final answer for the jpdf is therefore as claimed in eq.\ (\ref{PM}).
The normalisation constant can be determined by keeping track of all volume
factors and $\mu$-dependencies; it is given by
\bea
\fl
c_N=(VO(N)2^{-N}(2\pi)^{-N(N+1)/4})^2(2\pi)^{-N\nu/2}\,
\frac{VO(N+\nu)}{VO(N)VO(\nu)}
(2\mu)^{-N(N+\nu)}\eta_+^{-N(N+\nu-1)/2}.\nn\\
\label{CN}
\eea


\sect{Finite- and large-$N$ density correlation functions}\label{dense}


\subsection{The kernel}

The kernel $\mcK_N(z_1,z_2)$ as it initially appears in eq.\ (\ref{RnPf}) is
\textit{defined} as follows \cite{Sommers2007}
\be
\mcK_N(z_1,z_2) \equiv
\sum_{k=1}^{N}\sum_{l=1}^{N}\mcA_{kl}^{-1}z_1^{k-1}z_2^{l-1}
\ee
where the matrix $\mcA$ of dressed moments is related to $\mcF$ by
\be
\mcA_{kl} \equiv \int d^2z_1\,d^2z_2\,\mcF(z_1,z_2)\,z_1^{k-1}z_2^{l-1}.
\ee
However, to evaluate the matrix $\mcA$ and its inverse directly is not trivial
in general. Fortunately, the kernel may also be derived from the expectation
value of the product of two characteristic polynomials \cite{APSo}
\be
H_N(\lambda,\gamma) \equiv \langle \det(\lambda-\mcD)\,\det(\gamma-\mcD^{T})
\rangle_N,
\ee
where $\mcD$ was given in eq.\ (\ref{Mdef}). An explicit form for
$H_N(\lambda,\gamma)$ was derived by the authors in \cite{APSo}.\footnote{Note
  that $P$ and $Q$ were called $A$ and $B$ in \cite{APSo}; we set $n$ in
  \cite{APSo} to unity here.}
In the case of the real Ginibre ensemble
the spectral density was known prior to its integrability
\cite{Edelman97} and thus was used for the determination of the kernel in
\cite{Sommers2007}.

To establish the relationship between $\mcK_N(z_1,z_2)$ and
$H_{N-2}(\lambda,\gamma)$, we first relate the latter to the complex
eigenvalue density $R_{1,N}^{\mathbb{C}}(z)$.
If we choose $\gamma=\lambda^*$ with
$\im \, \lambda^2>\im \, \gamma^2$ we have
\bea
\fl H_{N-2}(\lambda,\la^*)&=& \int_{\mathbb{C}}
d^2z_1\ldots \int_{\mathbb{C}} d^2z_{N-2}
P_{N-2}(z_1,\ldots,z_{N-2})(\lambda\la^*)^\nu
\prod_{j=1}^{N-2}(\lambda^2-z_j^2)(\la^{*\,2}-z_j^2) \nn\\
\fl &=& \frac{c_{N-2}}{c_{N}}
\frac{1}{\exp[-\eta_-(\la^2+\la^{2\,*})]
g(\la^2)^2(\la^2-\la^{2\,*})(-2i)\Theta(\im  \lambda^2)}\nn
\\
\fl&&\times
\int_{\mathbb{C}}
d^2z_1\ldots \int_{\mathbb{C}}
d^2z_{N-2}\tilde{P}_{N}(z_1,\ldots,z_{N-2},\la^2,\la^{2\,*}) \ ,
\label{preHK}
\eea
where $\tilde{P}_{N}$ indicates that this jpdf is conditioned that the last
two eigenvalues are complex conjugates and ordered. The last line in
eq.\ (\ref{preHK}) is thus nothing but the complex density
$R_{1,N}^{\mathbb{C}}(\la^2)$, obtained by
inserting a delta function into the partition function together with the
constraint that the last two eigenvalues are complex conjugates.
We thus arrive at the following relationship
\be
R_{1,N}^{\mathbb{C}}(z) = \frac{c_{N}}{c_{N-2}}
\frac{w(z)w(z^*)}{|z|^\nu}\,(-2i)(z-z^*)\, H_{N-2}(\sqrt{z},\sqrt{z^*}).
\ee
On the other hand  we have from eq.\ (\ref{density}) an equation
that relates the complex
density directly to the kernel
\be
R_{1,N}(z) = \int_{\mathbb{C}} d^2u \, \mcK_N(z,u)\,\mcF(u,z)
\ee
and so for complex $z$ inserting eq.\ (\ref{Fdef})
\be
R_{1,N}^{\mathbb{C}}(z) = \mcK_N(z, z^*)\,w(z)w(z^*) (-2i)\ \sgn(y).
\ee
We can therefore make the identification (after analytic continuation in each
argument)
\be
\mcK_N(u,v) =
\frac{c_N}{c_{N-2}}(u-v)\,\frac{H_{N-2}(\sqrt{u},\sqrt{v})}{(uv)^{\nu/2}}.
\ee
Using the solution for $H_N(\lambda,\gamma)$ obtained in \cite{APSo}
as well as eq.\ (\ref{CN}), we then have for the properly normalised kernel
that
\bea
&&\fl
\mcK_N(u,v)
=  \frac{\eta_-}{8\pi(4\mu^2\eta_+)^{\nu+1}}
\sum_{j=0}^{N-2} \left( \frac{\eta_-}{\eta_+}
\right)^{2j}\frac{(j+1)!}{(j+\nu)!}   \left\{ L_{j+1}^{\nu} \left(
\frac{v}{4\mu^2\eta_-} \right) L_{j}^{\nu} \left( \frac{u}{4\mu^2\eta_-}
\right)
\right.\nn\\
&&\fl\quad\quad\quad\quad\quad\quad
\quad\quad\quad\quad\quad\quad\quad\quad\quad \quad\quad\quad \quad\quad
 \left. -L_{j+1}^{\nu} \left(
\frac{u}{4\mu^2\eta_-} \right) L_{j}^{\nu} \left( \frac{v}{4\mu^2\eta_-}
\right)\right\},
\label{KfiniteN}
\eea
which is the same as eq.\ (\ref{Kdef}). It can be further simplified to be
expressed as an anti-symmetric derivative as follows.
For modified Laguerre polynomials, we can use a recurrence relation to
show that
\bea
\fl
(j+1)\left\{ L_{j+1}^{\nu}(y)L_{j}^{\nu}(x) - ( x \leftrightarrow y ) \right\}
& = & \left( y \frac{\partial}{\partial y}L_{j}^{\nu}(y) +
(j+\nu+1-y)L_{j}^{\nu}(y) \right) L_{j}^{\nu}(x)\nn\\
\fl&& - (x \leftrightarrow y)  \nn \\
\fl
 & = & \left(y \frac{\partial}{\partial y} - x \frac{\partial}{\partial x} -
(y-x)\right)L_{j}^{\nu}(x)L_{j}^{\nu}(y).
\eea
The kernel therefore becomes
\bea\fl
\mcK_N(z_1,z_2) =
\frac{\eta_-}{8\pi(4\mu^2\eta_+)^{\nu+1}}
\left(y
\frac{\partial}{\partial y} - x \frac{\partial}{\partial x} - (y-x)\right)
\sum_{j=0}^{N-2} \left(
\frac{\eta_-}{\eta_+} \right)^{2j}\! \frac{j!}{(j+\nu)!}
L_{j}^{\nu}(x)L_{j}^{\nu}(y)
\nn\\
\label{Kdiff}
\eea
where $x$ and $y$ are evaluated at $x=z_1/4\mu^2\eta_-$ and
$y=z_2/4\mu^2\eta_-$ after the differentiations.
The symmetric kernel in terms of Laguerre polynomials
on the right hand side of eq.\ (\ref{Kdiff}) is
nothing else but the kernel of the
complex ($\beta=2$) two-matrix model \cite{Osborn}.

We have explicitly checked that a similar relation holds relating the kernel of
the non-chiral real Ginibre ensemble (see the equation after 5.26 in second
ref.\ \cite{Forrester07}) to $(\partial_y-\partial_x-2(y-x))$ operating on the
kernel of the non-chiral complex Ginibre ensemble (see eq.\ 40 in \cite{FKS})
and thus this is a more general feature.

Note that the convention used in \cite{APSo} for defining the `kernel' was
different from that adopted here (in eq.\ (\ref{KfiniteN})); in that paper,
there was no division by $(uv)^{\nu/2}$, and the summation ran to $N$. There
was also a minor 
typographical error in the arguments of one function (eq.\ 33 in
\cite{APSo}).


\subsection{Finite-$N$ results}

From eq.\ (\ref{RnPf}) we have
\be
R_1(z_1)= \int_{\mathbb{C}}  d^2z \, {\cal K}_N(z_1,z){\cal F}(z,z_1)
\equiv R_1^{\mathbb{C}}(z_1)+\delta(y_1)R_1^{\mathbb{R}}(x_1)\ .
\ee
We now simply insert the finite-$N$ kernel eq.\ (\ref{KfiniteN}) and weight
function  eq.\ (\ref{weight})  into this using eq.\ (\ref{Fdef}), to give
\bea
\fl R_1^{\mathbb{C}}(z)&=&  -2i \ \sgn(\Im m\, z) \
2\int_0^\infty \frac{dt}{t}\e^{-\eta_+^2t(z^2+z^{*\,2})-\frac{1}{4t}}
K_{\frac{\nu}{2}}\left(2\eta_+^2t|z|^2\right)
\erfc\left(2\eta_+\sqrt{t}|\Im m\,z|\right)
\nn\\
\fl&&\times
\frac{\eta_-|z|^{\nu}\e^{2\eta_-\Re e\,z}}{8\pi(4\mu^2\eta_+)^{\nu+1}}
\sum_{j=0}^{N-2} \left( \frac{\eta_-}{\eta_+}
\right)^{2j} \, \frac{(j+1)!}{(j+\nu)!}   \left\{ L_{j+1}^{\nu} \left(
\frac{z^*}{4\mu^2\eta_-} \right) L_{j}^{\nu} \left( \frac{z}{4\mu^2\eta_-}
\right) - \textup{c.c.} \right\}
\nn\\
\fl&&\label{rhoCN}
\\
\fl R_1^{\mathbb{R}}(x)&=&
\frac{\eta_-}{8\pi(4\mu^2\eta_+)^{\nu+1}}
\int_{-\infty}^\infty dx' \sgn(x-x')
|xx'|^{\nu/2}\e^{\eta_- (x+x')}2K_{\frac{\nu}{2}}(\eta_+|x|)
2K_{\frac{\nu}{2}}(\eta_+|x'|)  \nn \\
\fl&&\times\sum_{j=0}^{N-2} \left( \frac{\eta_-}{\eta_+} \right)^{2j} \,
\frac{(j+1)!}{(j+\nu)!}   \left\{ L_{j+1}^{\nu} \left( \frac{x'}{4\mu^2\eta_-}
\right) L_{j}^{\nu} \left( \frac{x}{4\mu^2\eta_-} \right) - (x'
\leftrightarrow x) \right\}.
\label{rhoRN}
\eea
These results are valid for even $N$ only. Alternatively we could have used
the form eq.\ (\ref{Kdiff}) which is more 
reminiscent of the corresponding chGOE result \cite{JacNc2} at $\mu=0$.
In fact we have checked that in the limit $\mu\to0$ the complex density eq.
(\ref{rhoCN}) vanishes, and the real density  eq. (\ref{rhoRN}) reduces to the
finite-$N$ expression (equation 5.18 in  \cite{JacNc2})
for the chGOE, after using
some identities for modified Laguerre polynomials.
We have also checked these results numerically using Monte Carlo, by
generating random matrices
and explicitly diagonalising them.

\begin{figure}[h]
  \unitlength1.0cm
\centerline{
\epsfig{file=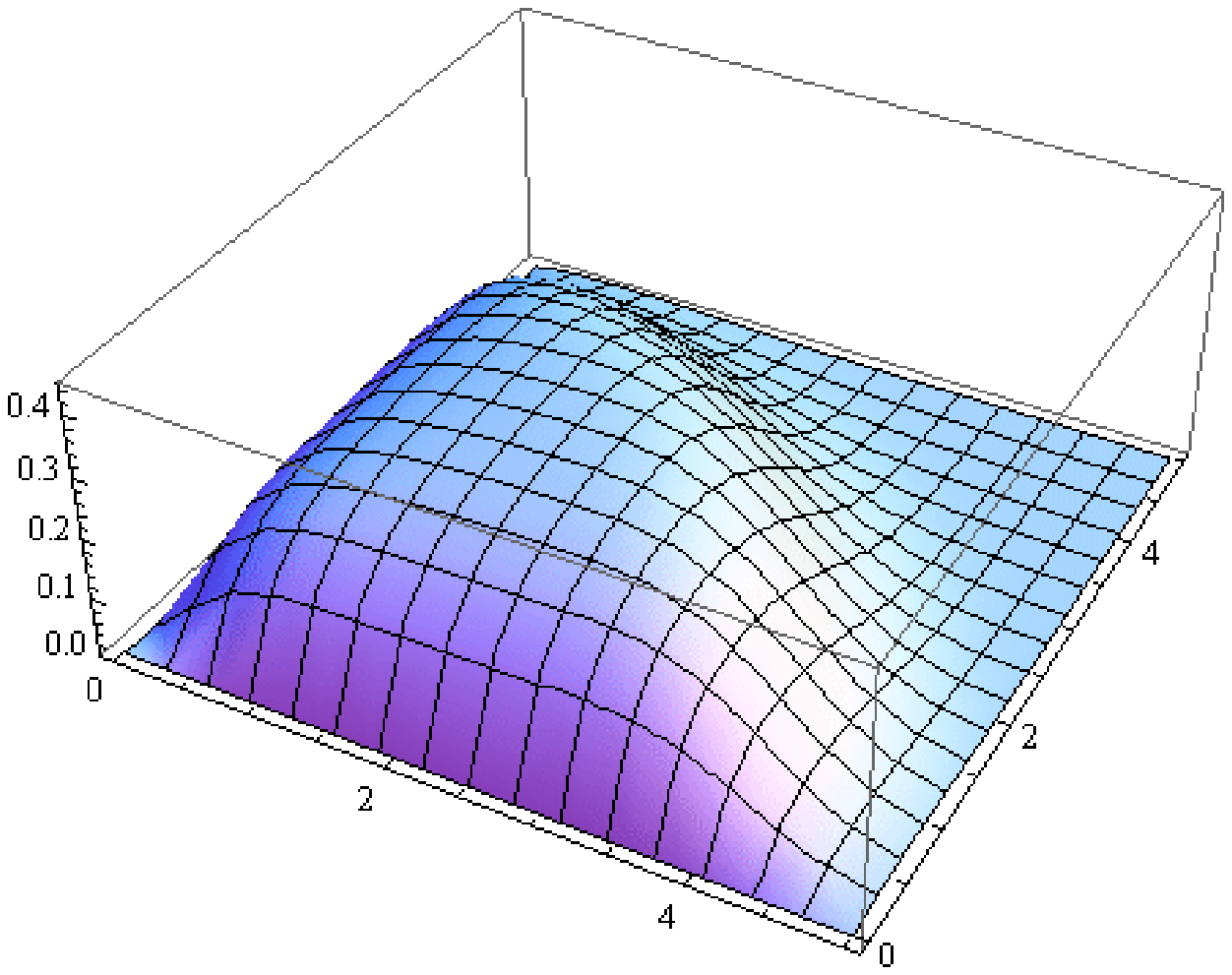,clip=,width=5.8cm}
\epsfig{file=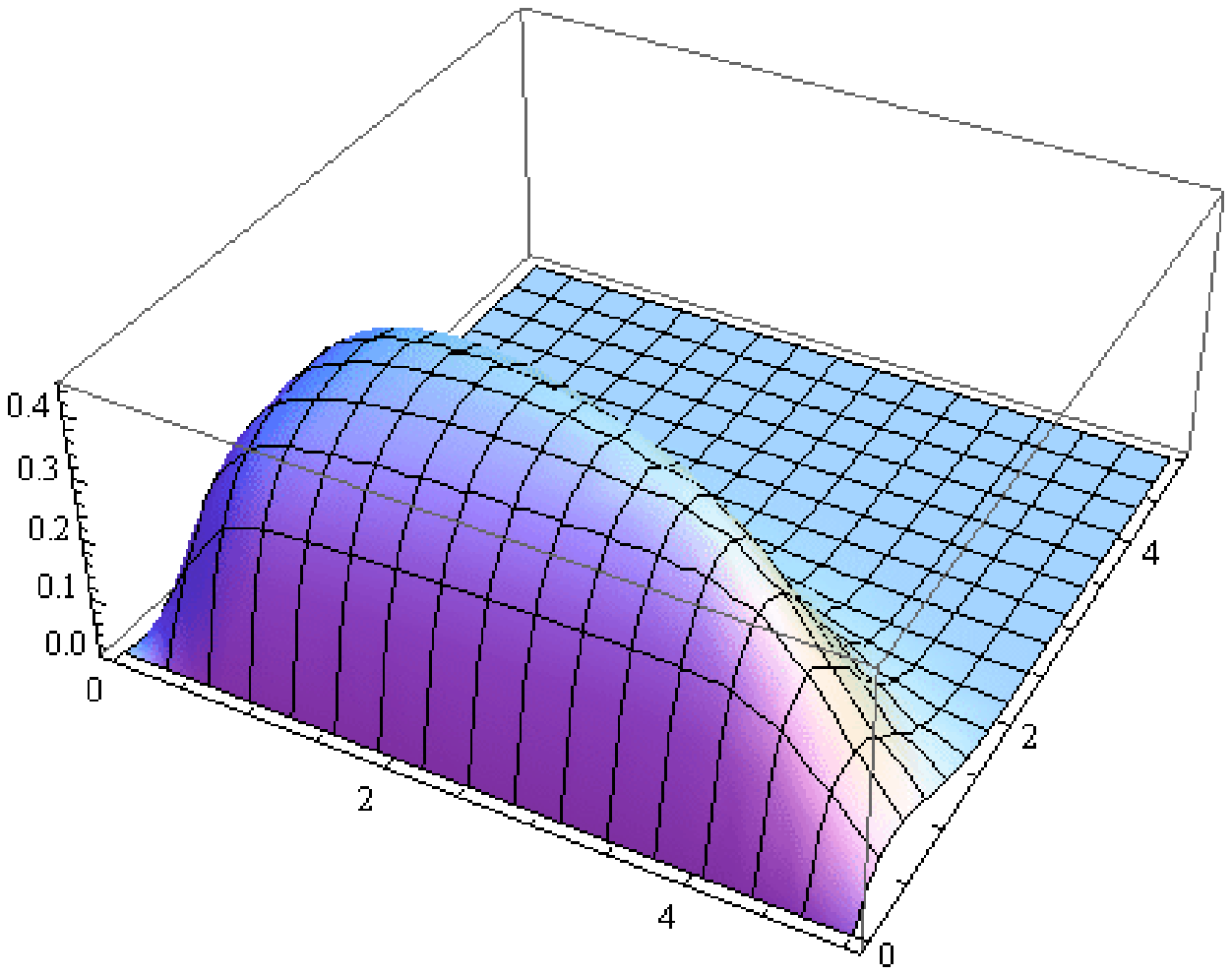,clip=,width=5.8cm}
\put(-3.7,0.2){$x$}
\put(-9.7,0.2){$x$}
\put(-0.5,1.2){$y$}
\put(-6.5,1.2){$y$}
\put(-5.8,4.2){$R_{1\,Dirac}^{\mathbb{C}}(z)$}
\put(-11.8,4.2){$R_{1\,Dirac}^{\mathbb{C}}(z)$}
}
  \caption{\label{rhoCstrfigN}
The complex spectral density $ R_{1\,Dirac}^{\mathbb{C}}(z)$ for finite $N=10$
at maximal non-Hermiticity $\mu^2=1$
(left) and intermediate $\mu^2=0.5$ (right), both for $\nu=0$.
We show only the first quadrant for symmetry reasons.
For $\mu=1$ we see a circular ``support''
growing with $\sqrt{N}$, apart from the
repulsion from the axes.
For decreasing $\mu$ the ``support'' becomes an ellipse, with the eigenvalues
moving towards, as well as onto, the real axis. Note the increased height in
the 
right plot.}
\end{figure}

\begin{figure}[h]
  \unitlength1.0cm
\centerline{
\epsfig{file=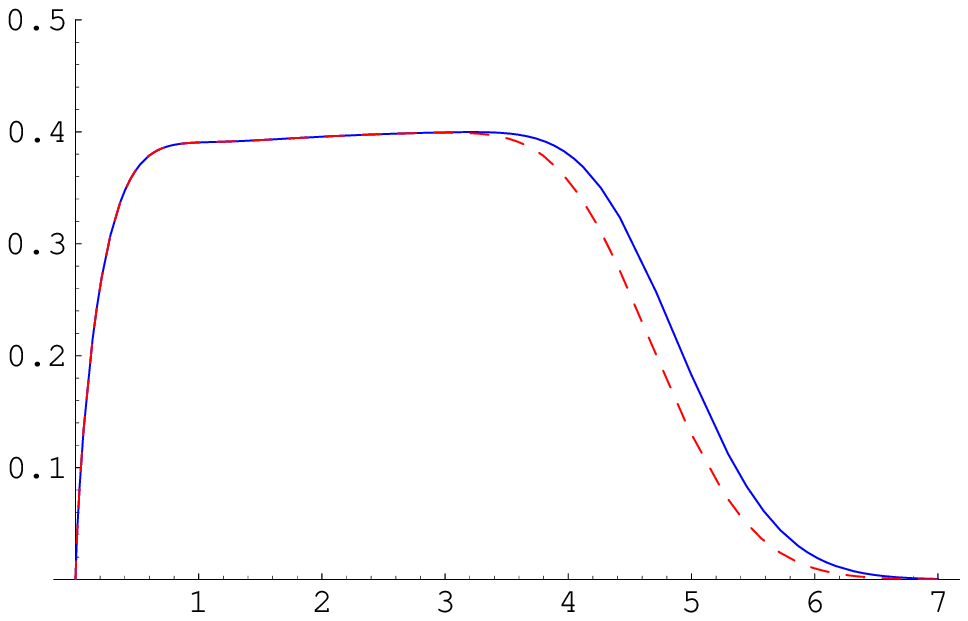,clip=,width=5.8cm}
\epsfig{file=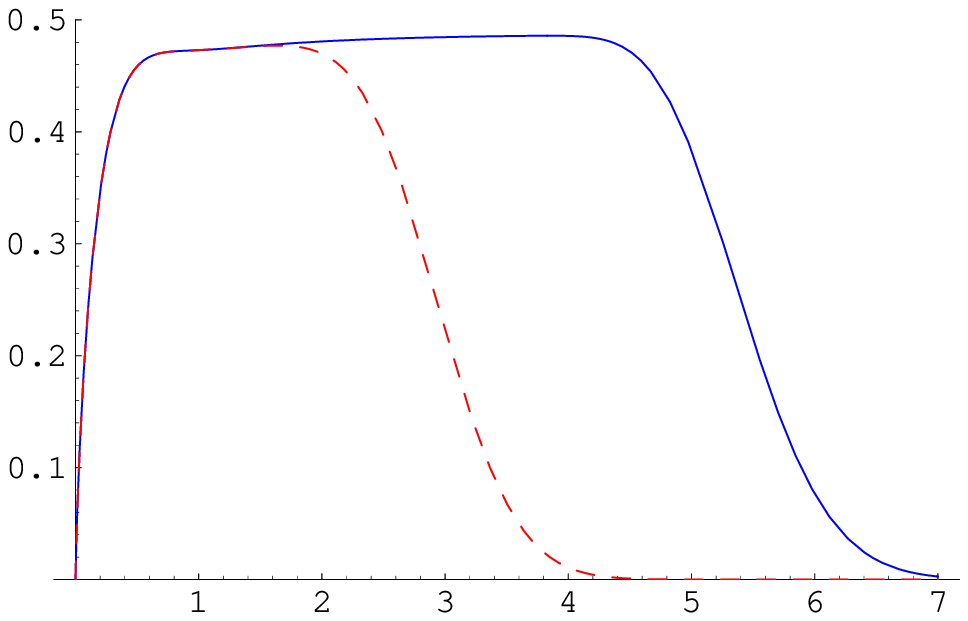,clip=,width=5.8cm}
\put(0.1,0.2){$x,$}
\put(-5.9,0.2){$x,$}
\put(-5.8,4.2){$R_{1\,Dirac}^{\mathbb{R}}((i)x)$}
\put(-11.8,4.2){$R_{1\,Dirac}^{\mathbb{R}}((i)x)$}
}
  \caption{\label{rhoRstrfigN}
The spectral density $ R_{1\,Dirac}^{\mathbb{R}}(x)$ for real eigenvalues
(blue full line)
and $ R_{1\,Dirac}^{\mathbb{R}}(ix)$ for purely imaginary eigenvalues
(red dashed line) for finite $N=10$
at almost maximal non-Hermiticity $\mu^2=0.95$ (left)
and intermediate $\mu^2=0.5$ (right), both for $\nu=0$. Here we have chosen
(almost) the same parameter values as Fig. \ref{rhoCstrfigN}, with
$\mu^2=0.95$ here close to 1 there.
Because of chiral symmetry real and imaginary
eigenvalues come in $\pm$ pairs and we only have to show the positive axes,
comparing both distributions in the same
plot. Whilst for $\mu^2=0.95$ the distributions of real and imaginary Dirac
eigenvalues are almost the same, for
$\mu^2=0.5$ there are more eigenvalues on the real than on the imaginary axis.
}
\end{figure}

As the last step we can change from squared variables  $z=x+iy=\La^2$ to Dirac
eigenvalues $\La$, using eq.\ (\ref{densitytrans}).
These two densities are illustrated\footnote{Here and in the following
  numerical integrals are carried out using \cite{Mathematica}.} 
in Figures \ref{rhoCstrfigN} and
\ref{rhoRstrfigN}, showing the localisation of the support for finite $N$.

\subsection{The large-$N$ limit at strong non-Hermiticity}

In the strong non-Hermitian limit, we keep $\mu$ fixed as we take the
large-$N$ limit.
This necessitates {\it no} rescaling of the eigenvalues (see
Section \ref{sum}), which is why our result is also true away from the origin.
Let us first determine the large-$N$ limit of the
kernel.

Reference \cite{Grad} eq.\ 8.976.1 gives
us the so-called Hille-Hardy formula (the equivalent of Mehler's formula
for Hermite polynomials):
\be\fl
S(x,y,z) \equiv \sum_{j=0}^{\infty} \frac{j!}{(j+\nu)!}
L_j^{\nu}(x)L_j^{\nu}(y)z^j =
\frac{(xyz)^{-\nu/2}}{1-z}\e^{-\frac{z}{1-z}\,(x+y)}I_{\nu}\left(
\frac{2\sqrt{xyz}}{1-z} \right).
\ee
We can insert this into  eq.\ (\ref{Kdiff}) and evaluate it after extending the
sum to infinity.
On differentiating the right-hand side, certain terms cancel, and we can
therefore establish that
\be
y\,\frac{\partial S(x,y,z)}{\partial y}
- x\, \frac{\partial S(x,y,z)}{\partial x} = - \,
\frac{z}{1-z}\,(y-x)\,S(x,y,z).
\ee
Hence the limit as $N \rightarrow \infty$ of the kernel is easily seen to be
\be\fl
\mcK^S_\nu(z_1,z_2) \equiv \lim_{N \rightarrow \infty} \mcK_N(z_1,z_2) =
\frac{\eta_+^3}{8\pi}
(z_1 - z_2)\,\e^{-\eta_-(z_1+z_2)}\,(z_1 z_2)^{-\nu/2}\,I_{\nu}\left(
2\eta_+\sqrt{z_1 z_2} \right).
\label{Kstrong2}
\ee
Because of this simplification this is now {\it proportional} to the $\beta=2$
kernel at strong non-Hermiticity.
Multiplication by the weight function eq.\ (\ref{weight}) which contains the
modulus $|z_1 z_2|^{\nu/2}$ will only cancel the pre-factor in
eq.\ (\ref{Kstrong2}) up to a phase.
Putting all ingredients together we can determine
the eigenvalue densities using eq.\ (\ref{density}):
\newpage
\bea
\fl
\rho_\nu^{\mathbb{C}\,S}(z) & = & \sgn(\Im m\, z)
(-2i)(z-z^*)\frac{\eta_+^3}{8\pi}
I_{\nu}\left( 2\eta_+|z| \right)
\label{rhoCstr}\\
\fl
 && \times 2\int_{0}^{\infty} \frac{dt}{t} \,
\exp\left[ -\eta_+^2(z^2+z^{*\,2})t -
\frac{1}{4t} \right] \,
K_{\frac{\nu}{2}}\left(2\eta_+^2|z|^2t\right) \, \erfc\left(2\eta_+
\sqrt{t}\,|\Im m\, z|\right),
\nn\\
\fl
\rho_\nu^{\mathbb{R}\,S}(x) & = & \frac{\eta_+^3}{8\pi}
 \, 2K_{\frac{\nu}{2}}(\eta_+|x|) \left(
\int_{0}^{\infty}dx'\,|x -x'|\
2K_{\frac{\nu}{2}}(\eta_+|x'|)\
I_{\nu}( 2\eta_+\sqrt{x x'})\right.
\label{rhoRstr}\\
\fl&&\quad\quad\quad\quad\quad\quad\left.
+\int_{-\infty}^{0}dx'\,|x -x'|\
2K_{\frac{\nu}{2}}(\eta_+|x'|)\
J_{\nu}( 2\eta_+\sqrt{x |x'|})
\right).
\nn
\eea
Note the change from Bessel-$I$ to Bessel-$J$ function inside the integral for
negative arguments, after taking into account the aforementioned phase.

If we rescale the eigenvalues as $2\eta_+z^2\to z^2$ (and divide the densities
by $2\eta_+$ accordingly) we obtain the same densities as at maximal
non-Hermiticity. These are obtained by setting $\mu=1\Rightarrow\eta_+=\frac12$
above, or by starting from the kernel eq.\ (\ref{Kstr}) at maximal
non-Hermiticity. This feature that the strong limit can be obtained by
rescaling the case of maximal  non-Hermiticity is generically true for complex
RMT, see \cite{FKS98}.
\begin{figure}[h]
  \unitlength1.0cm
\centerline{
\epsfig{file=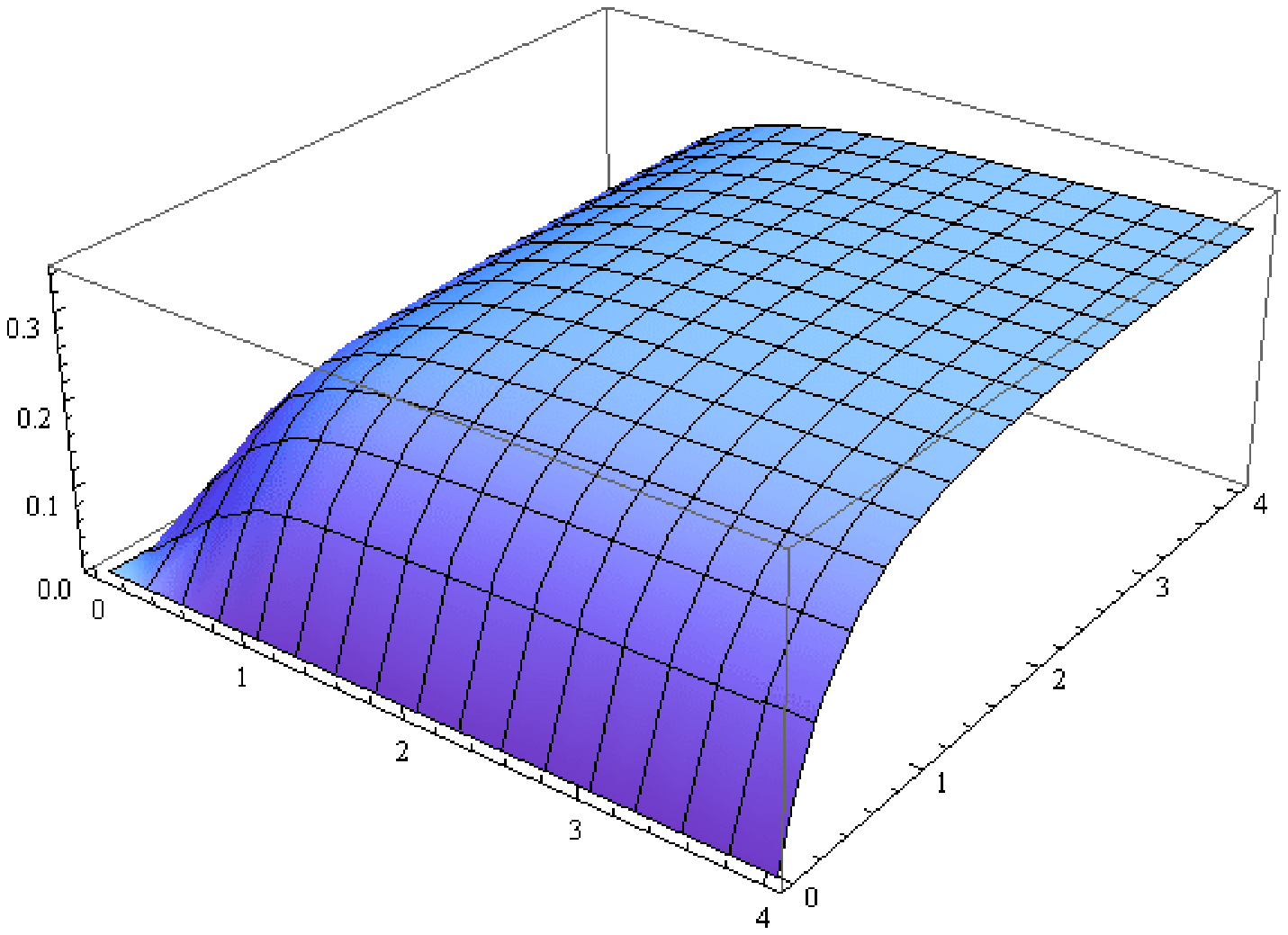,clip=,width=5.8cm}
\epsfig{file=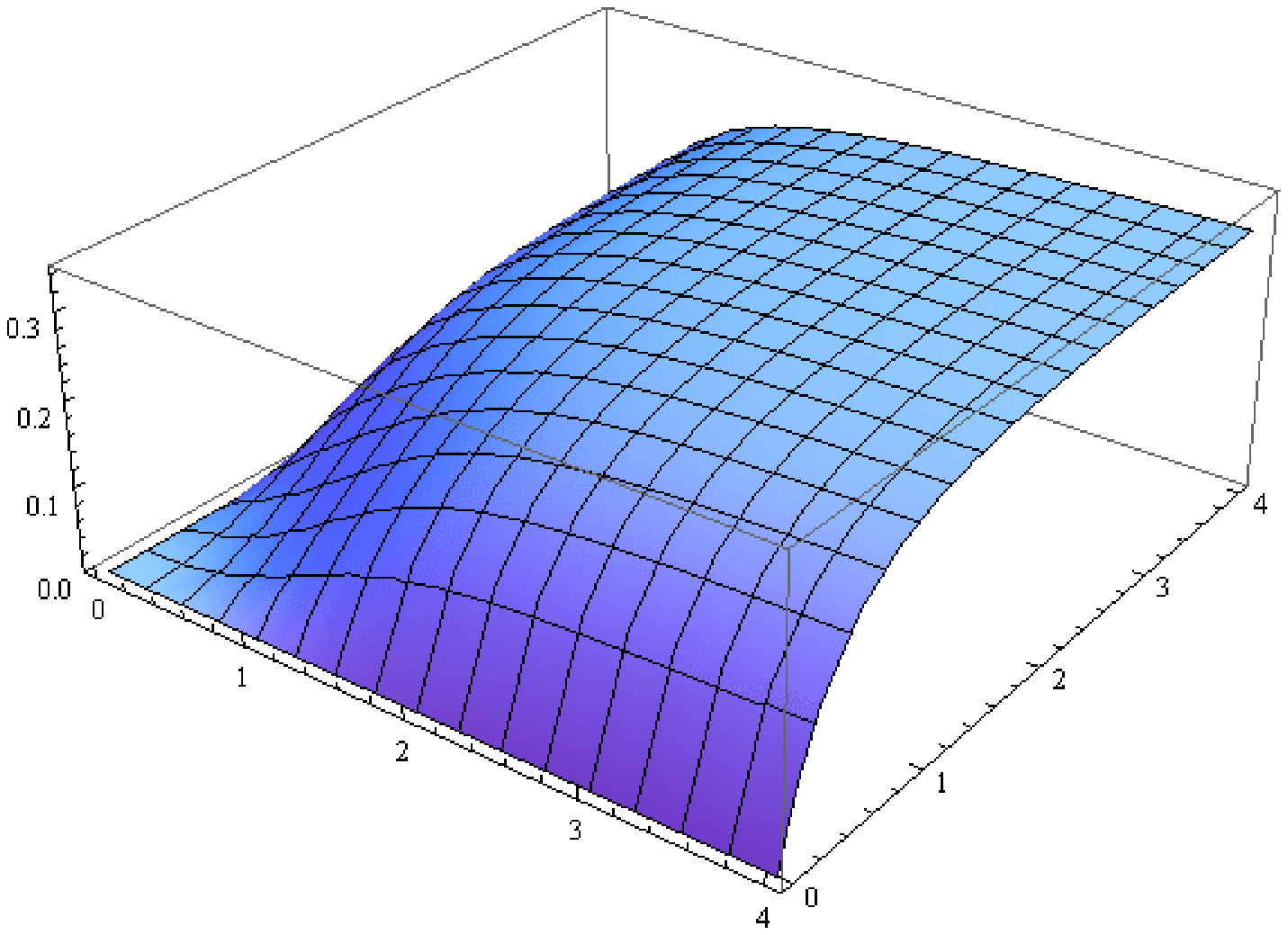,clip=,width=5.8cm}
\put(-3.7,0.2){$x$}
\put(-9.7,0.2){$x$}
\put(-0.5,1.2){$y$}
\put(-6.5,1.2){$y$}
\put(-5.8,4.2){$\rho_{2\, Dirac}^{\mathbb{C}\,S}(\La)$}
\put(-11.8,4.2){$\rho_{0\, Dirac}^{\mathbb{C}\,S}(\La)$}
}
  \caption{\label{rhoCstrfig}
The complex spectral density $\rho_{\nu\ Dirac}^{\mathbb{C}\,S}(\La=x+iy)$
at maximal non-Hermiticity $\mu=1$. It is
shown only in the
first quadrant for symmetry reasons, for $\nu=0$
(left) and $\nu=2$ (right). Increasing the number of exact zero eigenvalues
$\nu$ leads to a stronger repulsion from the origin. At $\nu=0$ this repulsion
is still present due to chiral symmetry (or technically speaking
the presence of the Bessel-$K$ function).}
\end{figure}

\begin{figure}[h]
  \unitlength1.0cm
\centerline{
\epsfig{file=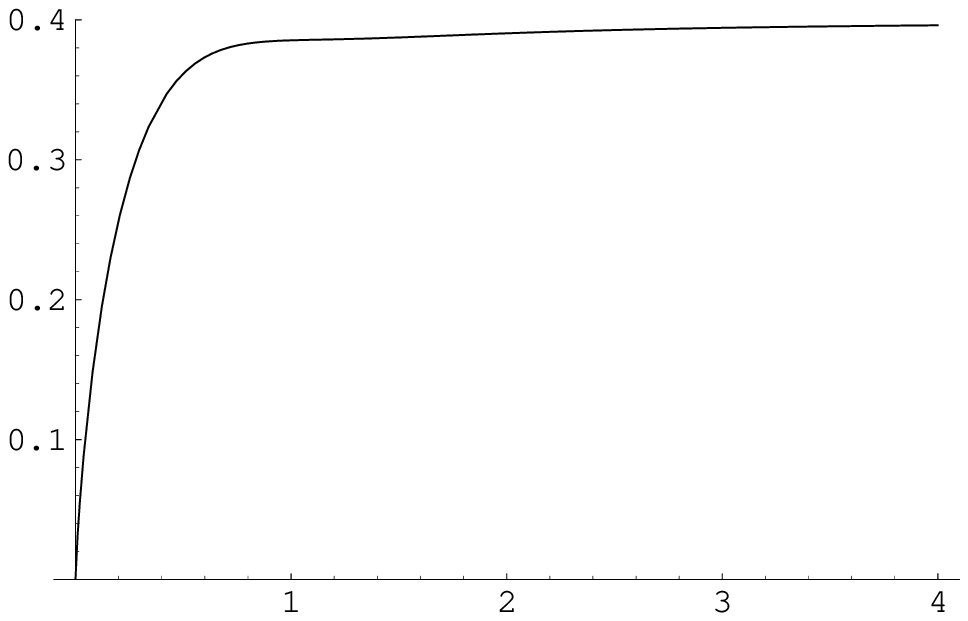,clip=,width=5.8cm}
\epsfig{file=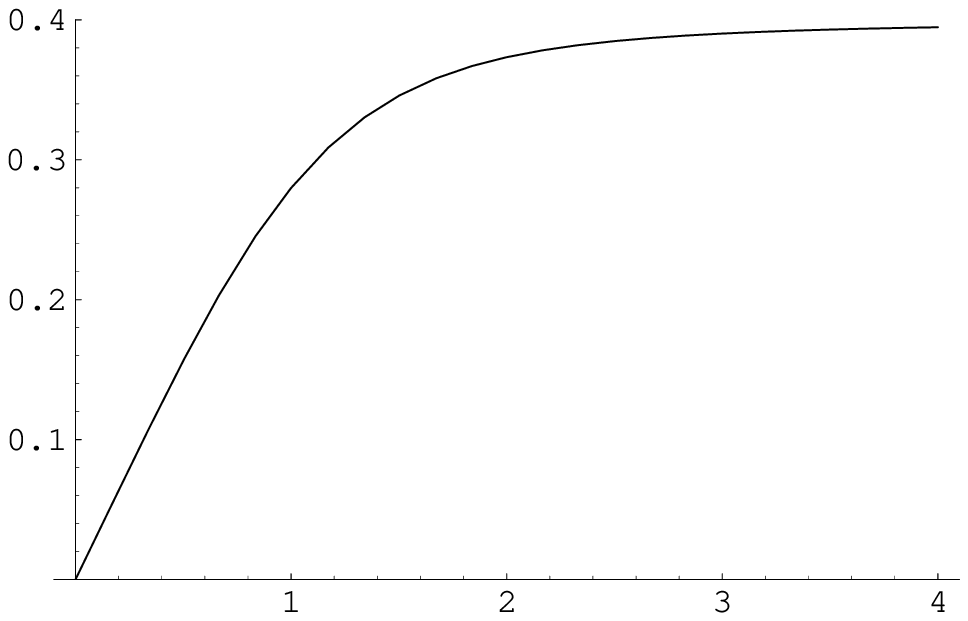,clip=,width=5.8cm}
\put(-0.0,0.2){$x$}
\put(-6.0,0.2){$x$}
\put(-5.8,4.2){$\rho_{2\, Dirac}^{\mathbb{R}\,S}(x)$}
\put(-11.8,4.2){$\rho_{0\, Dirac}^{\mathbb{R}\,S}(x)$}}
  \caption{\label{rhoRstrfig}
The real spectral density of Dirac eigenvalues on the positive half-line
at maximal non-Hermiticity $\mu=1$ for $\nu=0$
(left) and $\nu=2$ (right). Because of chiral symmetry it is
symmetric on the negative real line, and because $\mu=1$ it is identical on
the imaginary axis. For $\nu=2$ we see the increased repulsion from the origin
compared to $\nu=0$, as for the complex eigenvalues in Fig. \ref{rhoCstrfig}.}
\end{figure}

In Figures \ref{rhoCstrfig} and \ref{rhoRstrfig}
we show the densities of Dirac eigenvalues $\La$ for complex eigenvalues
and real or purely imaginary eigenvalues
respectively,
using the mapping eq.\ (\ref{densitytrans}). Because of the rescaling property
just mentioned
we only show results here for maximal non-Hermiticity.
One can check analytically using a
saddle-point approximation including the fluctuations that for asymptotically
large $x,y$ the densities eqs.\ (\ref{rhoCstr}) and (\ref{rhoRstr}) decay as
$\sim1/|z|$ and $1/\sqrt{x}$ respectively. After the mapping to Dirac
eigenvalues eq.\ (\ref{densitytrans}) they thus reach a plateau as seen in
the figures.
We note that the profiles of the densities on the real and imaginary axis 
are very reminiscent to parallel cuts through the complex densities, for both
values of $\nu$ shown.

\subsection{The large-$N$ limit at weak non-Hermiticity}

In the weak case, we scale $\mu$ and $z$ with $N$ as follows (as compared with
the unscaled case, i.e. we implicitly include
here the first rescaling discussed in Section \ref{N})
\be
\mu \equiv \frac{\alpha}{\sqrt{2N}}\ ,\ \ 4Nz\equiv \hat{z}
\label{alscale}
\ee
where $\alpha$ and $\hat{z}$ are
kept fixed throughout. Because of the rescaling of the eigenvalues here we
are magnifying the region around the origin.

For the weight function we thus simply obtain
\be\fl
\lim_{N\rightarrow\infty}
K_{\frac{\nu}{2}}\left(\eta_+\,\frac{|z|}{4N}\right) = K_{\frac{\nu}{2}}
\left( \frac{|z|}{8\alpha^2} \right) \quad \textup{and} \quad
\lim_{N\rightarrow\infty} \exp\left[\eta_-\frac{z}{4N}\right] =
\exp\left[\frac{z}{8\alpha^2}\right].
\label{weakweight}
\ee
For the kernel we are interested in the limit of
eq.\ (\ref{KfiniteN}) in terms of the rescaled variables, i.e.
$\mu$, $\eta_-$ and $\eta_+$ which all now depend on $N$.
Instead of using eq.\ (\ref{Kdiff}) it is slightly
simpler if we can rewrite eq.\ (\ref{KfiniteN}) so that the Laguerre
polynomials
inside the sum are of the same degree $j$.  Using the recurrence relationship
for the Laguerre polynomials (eq.\ 8.971.2 in \cite{Grad}), we have
\bea\fl
(n+1)\left\{ L_{n+1}^{\nu}(v)\,L_{n}^{\nu}(u) - (u \leftrightarrow v) \right\}
& = & \left((n+1+\nu)L_{n}^{\nu}(v) - vL_{n}^{\nu+1}(v)\right)\,L_{n}^{\nu}(u)
- (u \leftrightarrow v)  \nn \\
 & = & uL_n^{\nu+1}(u)\,L_n^{\nu}(v) - vL_n^{\nu+1}(v)\,L_n^{\nu}(u)\ .
\eea
The kernel can therefore be written as
\bea
\fl
\mcK_N\left({z_1},{z_2}\right) &
=&
\frac{1}{8\pi(4\mu^2\eta_+)^{\nu+1}4\mu^2}
\sum_{j=0}^{N- 2} \left( \frac{\eta_-}{\eta_+}
\right)^{2j} \frac{j!}{(j+\nu)!}
\label{kscale}\\
\fl && \times \left\{ z_1L_{j}^{\nu+1}\left(
\frac{z_1}{4\mu^2\eta_-} \right)L_{j}^{\nu}\left( \frac{z_2}{4\mu^2\eta_-}
\right) -
 z_2L_{j}^{\nu+1}\left(
\frac{z_2}{4\mu^2\eta_-} \right)L_{j}^{\nu}\left( \frac{z_1}{4\mu^2\eta_-}
\right)
\right\}.
\nn
\eea
We now wish to take the limit $N \rightarrow \infty$.
For this to exist we must multiply the kernel by the spacing as well as by
the appropriate number of zero-eigenvalues from the weight, as given in
eq.\ (\ref{Kweak2}) below.
In eq.\ (\ref{kscale}) we will replace
the sum with an integral over the variable $t \equiv \frac{j}{N} \in [0,1]$.
Because of the different scaling in the weak limit we cannot use the
Hille-Hardy formula as before.

In detail, using eq.\ 8.978.2 in \cite{Grad} we have for some real constant
$\nu$ and fixed $t \in [0,1]$ the standard Bessel asymptotic of the
modified  Laguerre polynomials:
\bea
\lim_{N \rightarrow \infty} \left[
  N^{-\nu}\,L_{tN}^{\nu}\left(\frac{x}{N}\right) \right] & = &
t^{\nu/2}x^{-\nu/2}\,J_{\nu}(2\sqrt{xt}).
\eea
We also have (keeping $t$ fixed so that $j=tN\to\infty$)
\be
\lim_{N \rightarrow \infty} \left(
\frac{1-\frac{\alpha^2}{2N}}{1+\frac{\alpha^2}{2N}} \right)^{2j} =
\exp[-2\alpha^2 t] \quad \textup{and} \quad \lim_{N \rightarrow \infty}
\frac{j!}{(j+\nu)!}\,N^{\nu} = t^{-\nu}.
\ee
Therefore,
\bea
\fl
\mcK^W(z_1, z_2) & \equiv & \lim_{N \rightarrow \infty} \left[
\frac{1}{(4N)^2}\left(\frac{z_1 z_2}{(4N)^2} \right)^{\nu/2}
  \mcK_{N}\left(\frac{z_1}{4N},\frac{z_2}{4N};\mu=\frac{\alpha}{\sqrt{2N}}
\right) \right] \label{Kweakdef} \\
\fl & = & \frac{1}{256 \pi\al^2} \int_0^1 \,ds\, s^2\,\e^{-2\alpha^2 s^2}\,
\left\{\sqrt{z_1}\,J_{\nu+1}(s\sqrt{z_1})J_{\nu}(s\sqrt{z_2}) - (z_1
\leftrightarrow z_2) \right\}
\label{Kweak2}\\
\fl&=& \frac{1}{128 \pi\al^2}\left(z_2\frac{\partial}{\partial z_2}
-z_1\frac{\partial}{\partial z_1}\right)
\int_0^1 ds\ s\
\e^{-2\al^2s^2}
J_\nu\left(s\sqrt{z_1}\right)J_\nu\left(s\sqrt{z_2}\right)
\nn
\eea
after a simple change of variables in the integral. In the last line we
expressed the kernel as a derivative of the $\beta=2$ kernel at weak
non-Hermiticity. This corresponds to eq.\ (\ref{Kdiff}) in which the second
term $-(y-x)$ now becomes 
sub-leading compared with the derivatives.
The same relation between the large-$N$ kernel at weak non-Hermiticity is true
for 
the $\beta=1$ and $\beta=2$ Ginibre ensembles \cite{Forrester07,FKS}
respectively. 

Collecting all elements we have for the complex density
\bea
\fl
\rho_\nu^{\mathbb{C}\,W}(z) & =&
-2i\ \sgn(\im \,z)\exp\left[\frac{1}{8\al^2}2\,\re\,z\right]
{\cal K}^W(z,z^*) \label{RcomplW}\\
\fl&&\times
2\,\int_0^\infty
\frac{dt}{t}\exp\Big[-\frac{t}{64\al^4}(z^2+z^{*\,2})-\frac{1}{4t}\Big]
K_{\frac{\nu}{2}}\left(\frac{t}{32\al^4}|z|^2\right)
\erfc\Big(\frac{\sqrt{t}}{4\al^2}|\im \, z|\Big).
\nn
\eea
\begin{figure}[h]
  \unitlength1.0cm
\centerline{
\epsfig{file=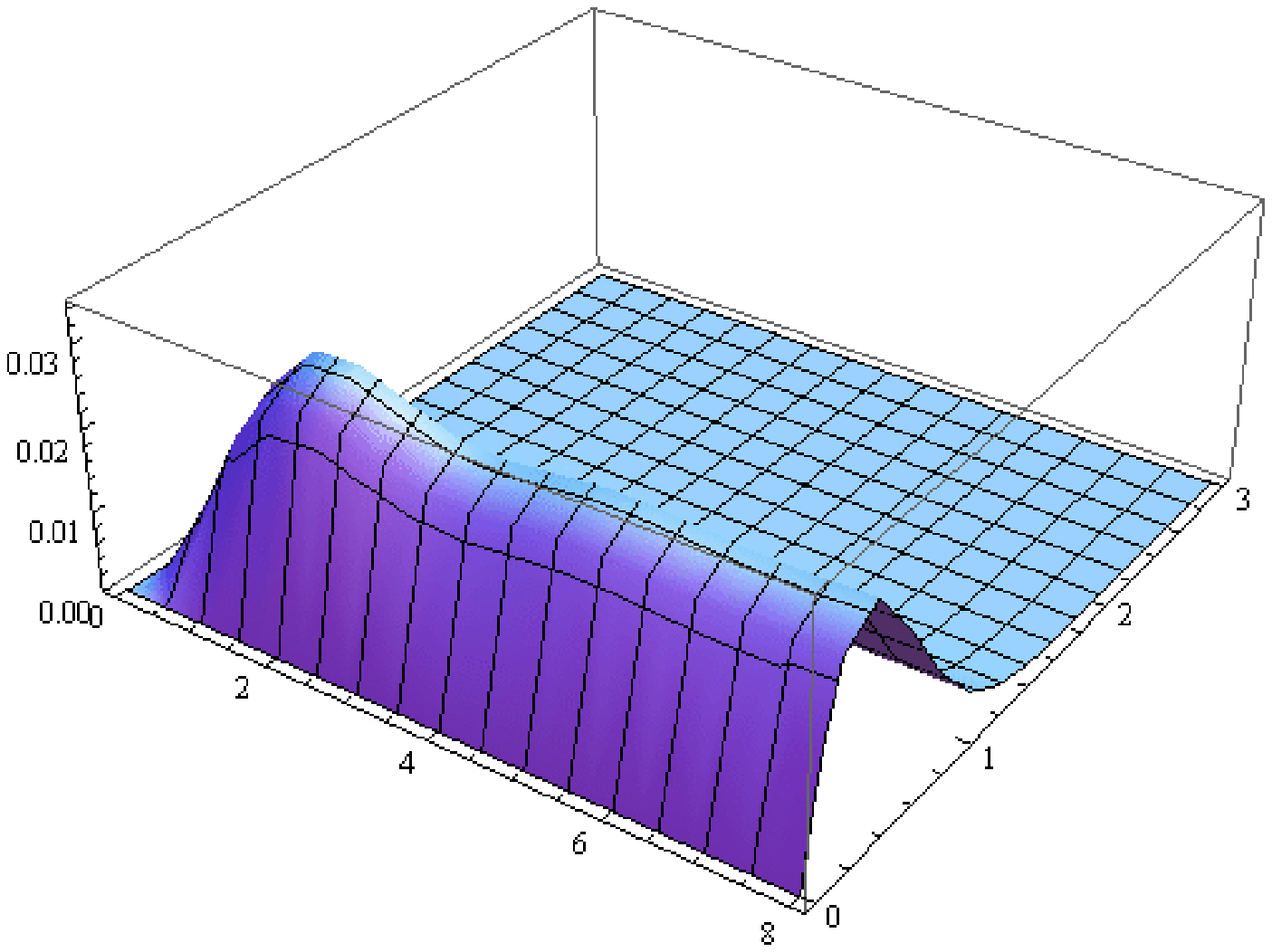,clip=,width=5.8cm}
\epsfig{file=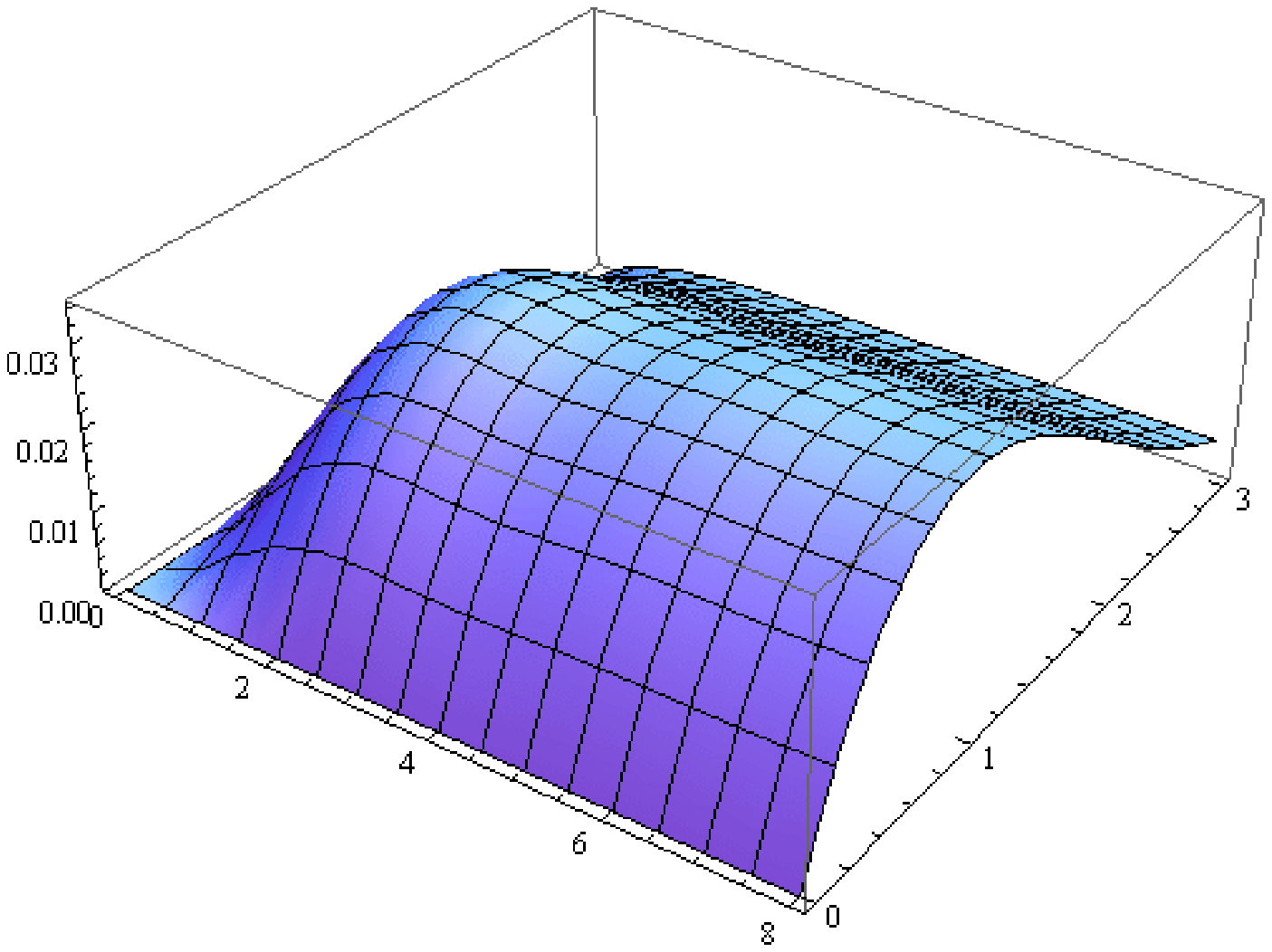,clip=,width=5.8cm}
\put(-3.7,0.2){$x$}
\put(-9.7,0.2){$x$}
\put(-0.5,1.2){$y$}
\put(-6.5,1.2){$y$}
\put(-5.8,4.2){$\rho_{0\, Dirac}^{\mathbb{C}\,W}(\La)$}
\put(-11.8,4.2){$\rho_{0\, Dirac}^{\mathbb{C}\,W}(\La)$}
}
  \caption{\label{rhoCweakfig}
The complex spectral density $\rho_{\nu\ Dirac}^{\mathbb{C}\,W}(\La=x+iy)$
at weak non-Hermiticity for parameters $\al^2=0.2$
(left) and $\al^2=1$ (right), both at $\nu=0$ plotted on the same scale.
For increasing $\al$
more complex eigenvalues move in towards the imaginary axis, reaching
Fig. \ref{rhoCstrfig} left in the limit $\al\to\infty$.}
\end{figure}

In Figure \ref{rhoCweakfig} it is shown after mapping to Dirac eigenvalues.
As a consistency check we can take the limit $\al\to\infty$
while keeping $z/\al^2$ fixed to obtain once more the complex density in the
strong 
non-Hermiticity limit, eq.\ (\ref{rhoCstr}).
The precise mapping of weak to strong eigenvalues
is given by $\frac{z}{4\al^2}\to2\eta_+z$. Whilst the matching of the integrals
over $t$ is straightforward the mapping of the kernels multiplied by the
weight is 
more involved. Changing variables we have the following identity:
\bea
&&\al^2\int_0^1 \,ds\, s^2\,\e^{-2\alpha^2 s^2}\,
\left\{\sqrt{z_1} J_{\nu+1}(s\sqrt{z_1})J_{\nu}(s\sqrt{z_2}) - (z_1
\leftrightarrow z_2) \right\}
\nn\\
&=& 2 \,
\left(z_2\frac{\partial}{\partial z_2}
-z_1\frac{\partial}{\partial z_1}\right)
\int_0^\al dt\ t\
\e^{-2t^2}
J_\nu\left(t\frac{\sqrt{z_1}}{\al}\right)
J_\nu\left(t\frac{\sqrt{z_2}}{\al}\right)\nn\\
&\to&
\frac{1}{2} \, \left(z_2\frac{\partial}{\partial z_2}
-z_1\frac{\partial}{\partial z_1}\right)
\exp\left[ -\,\frac{(z_1+z_2)}{8\al^2}\right]I_\nu\left(
\frac{\sqrt{z_1z_2}}{4\al^2}\right) \,
\label{intid}
\eea
where in the last step we have extended the integral to infinity and used
eq.\ 6.633.2 \cite{Grad}. The last differentiation is trivial, in effect
acting only on 
the exponential function. We thus obtain for the limiting kernel
\be
\fl
\lim_{\al\to\infty} \al^4 \mcK^W(z_1, z_2)\Big|_{\frac{z}{\alpha^2}
\textup{ fixed}} =
\left(\frac{z_1}{4\al^2} -\frac{z_2}{4\al^2}\right) \exp\left[
  -\,\frac{(z_1+z_2)}{8\al^2}\right]
I_\nu\left( \frac{\sqrt{z_1z_2}}{4\al^2}\right),
\ee
which precisely cancels the exponentials from the weight eq.\
(\ref{weakweight})
 to arrive at the complex density at strong non-Hermiticity
 eq.\ (\ref{rhoCstr}).

We now turn to the real density at weak non-Hermiticity. Looking at the
definitions eqs. (\ref{rhoCN}) and (\ref{rhoRN}) the main difference to the
complex density is that here the kernel is integrated, whereas the complex
density is simply given by the kernel multiplied by the weight. 

Unfortunately, and in contrast to the strong non-Hermitian case, 
at weak non-Hermiticity the large-$N$ limit and the
integration do {\it not} commute, with the integral over the weak kernel 
(\ref{Kweak2}) not being absolutely convergent. Such a feature 
might have been expected, as the same phenomenon occurs for 
the chGOE at $\mu=0$ \cite{JacNc2}. However, in that case, 
the integrals could be
computed exactly before taking the large-$N$ limit, leading to the correct
result, which differs from the naive limit by a factor of 2 in the
normalisation\footnote{A quick guess generalising this to our setting
fails.}. 
The integrals in eq. (\ref{rhoRN}) at finite $N$ are more involved, and this
matter will be addressed in future work.

\begin{figure}[h]
  \unitlength1.0cm
\centerline{
\epsfig{file=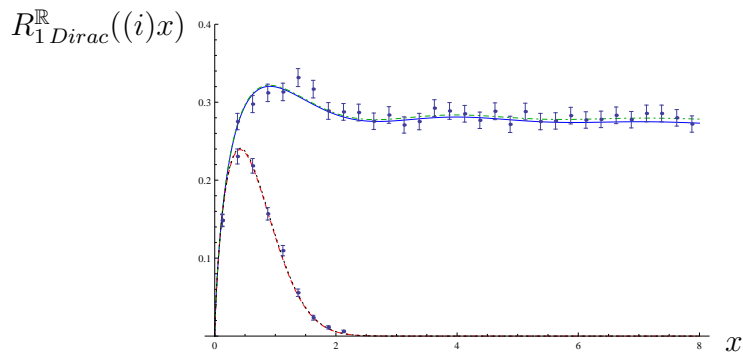,clip=,width=6.8cm}
\put(0.2,0.0){$x$}
\put(-9.3,4.2){$R_{1\, Dirac}^{\mathbb{R}}((i)x)$}
}
  \caption{\label{rhoRweakfig}
The real spectral density of Dirac eigenvalues on the positive half-line
scaled as for weak non-Hermiticity for $\al^2=\mu^2/2N=0.2$
at $\nu=0$. We show results for finite $N$ vs numerical simulations.
Both the eigenvalue densities for real eigenvalues 
$R_{1, Dirac}^{\mathbb{R}}(x)$ and for pure imaginary ones 
$R_{1, Dirac}^{\mathbb{R}}(ix)$ are displayed in the same plot for comparison.
Real density:
$N=10$ (blue full line), $N=20$ (dark green dot-dashed line); 
imaginary density $N=10$ (red dashed line)
$N=20$ (black, dotted line).   
Dots with error bars:  $N=100$ Monte Carlo simulation of $10^4$ matrices.
}
\end{figure}

For that reason we show in Fig. \ref{rhoRweakfig} the real 
density for finite but large $N=10$ and $20$, 
using the weak scaling from eq. (\ref{Kweakdef}) and
eq. (\ref{alscale}) for a given $\alpha$. 
This underlines that the weak limit for the real density does exist and
convergence is rapid. 
Furthermore we have checked this by superimposing data for
$N=100$ using numerically generated random matrices.

As a feature common to the strong limit we note that the densities of real and
purely imaginary eigenvalues in Fig. \ref{rhoRweakfig} resemble cuts through
the complex density in  Fig. \ref{rhoCweakfig} left at the same value of
$\alpha^2=0.2$.

\sect{Conclusions}\label{conc}

In this paper we have solved the chiral extension of the Ginibre ensemble of
real asymmetric matrices. It is given as a two-matrix model of rectangular
matrices with real elements and depends on a non-Hermiticity parameter $\mu$.
This model is relevant for computing the non-Hermitian spectrum
of Dirac operators with real elements in field theory. Our work completes the
programme of solving the three chiral or Wishart-Laguerre counterparts  of the
classical Ginibre ensembles, where earlier works  by Osborn and one of the
authors extended the models with
complex and quaternion real elements respectively.

Whilst our model inherits most
of the integrable structure of the real Ginibre ensemble its joint probability
distribution required a more complicated
calculation, which took much of our effort here.
Just as in the Ginibre ensembles the probability density for the matrix
elements is Gaussian, whereas the one for the eigenvalues becomes non-Gaussian.
It contains a Bessel-$K$ function and integral thereof, replacing the role of
the complementary error function in the real Ginibre ensemble.

The main building block for all eigenvalue correlation functions
is given by a kernel of Laguerre polynomials in the complex plane and was
derived in a previous paper. Here, we give all eigenvalue density correlation
functions for finite (even) $N$ valid for all values of $\mu$, in particular
the 
spectral one-point densities for real eigenvalues and for complex
non-real eigenvalues. 
Moreover, we have uncovered a way of expressing the $\beta=1$ kernel in terms
of 
the $\beta=2$ kernel at finite and large $N$,
both for the chiral and non-chiral Ginibre ensembles.
We conjecture that a similar relation holds for the $\beta=4$ kernel as well.

When taking microscopic
large-$N$ limits we focus on the origin where the chiral
symmetry of our model is the most important. For both the limit
at strong non-Hermiticity with
$0<\mu^2<1$, and the limit at weak
non-Hermiticity with $\mu^2\sim1/N$, we give compact expressions for the
kernel. This leads to explicit expressions for the complex spectral 
one-point densities and the real density at strong non-Hermiticity.

It would be very interesting to compare these results with simulations from
non-Hermitian lattice gauge theory, as was successfully done previously for
the other two chiral two-matrix models.

Further extensions would be to investigate the bulk or the soft edge scaling
limit. We expect that in the former the
chiral ensembles will agree with the Ginibre ensembles, as the effect of
chirality becomes unimportant in the bulk.
\\

This work has been supported partly by
European Network ENRAGE MRTN-CT-2004-005616 (G.A.),
an EPSRC doctoral training grant (M.J.P.) and the
SFB/TR12 of the Deutsche Forschungsgemeinschaft (H.-J.S.).
We thank Tilo Wettig for useful discussions.

\begin{appendix}
\sect{Details on the calculation of the Jacobian}\label{Jacobi}

In this appendix we give a few more details to complete the computation of the
Jacobian from Section \ref{Ngen}. In particular we first give a precise
ordering of matrix elements leading to a block-triangular Jacobian. Second we
will treat the case of non-diagonal matrices $\La_A$ and $\La_B$.

For the first purpose we repeat eq.\ (\ref{dAdB}) including matrix indices,
after dropping the outside rotations:
\bea
\fl
(dA)_{ij}\! & = & \sum_{k=1}^N (O_A^{T}dO_A)_{ik}(\Delta_A+\Lambda_A)_{kj} -
\!\sum_{k=1}^{N+\nu} (\Delta_A+\Lambda_A)_{ik}(O_B^{T}dO_B)_{kj}\nn\\
\fl&&+(d\Delta_A)_{ij} + (d\Lambda_A)_{ij}\ ,
\label{dA}\\
\fl
(dB^T)_{ij}\! & = &
\sum_{k=1}^{N+\nu}(O_B^{T}dO_B)_{ik}(\Delta_B+\Lambda_B)_{kj}
-\!\sum_{k=1}^N (\Delta_B+\Lambda_B)_{ik}(O_A^{T}dO_A)_{kj}\nn\\
\fl&&+(d\Delta_B)_{ij} + (d\Lambda_B)_{ij}\ .
\label{dB}
\eea
We now give an ordering leading to a block-triangular Jacobi matrix, with
variables $\{dA,dB^T\}$ in the columns and
$\{d\Lambda_A,d\Lambda_B,d\Delta_A,d\Delta_B,O_A^TdO_A,O_B^TdO_B\}$ in the
rows. For the block diagonal matrices $\La_{A,B}$ and upper block-diagonal
matrices $\Delta_{A,B}$ (of different size) this is trivial: we just group
them together with the corresponding elements of $dA$ and $dB^T$. For example,
when $N$ is even and $\nu >0$, this will give
\bea
\fl
&&(dA)_{11},(dA)_{12},(dA)_{21},(dA)_{22},(dA)_{33},\ldots,(dA)_{N\!N},
(dB^T)_{11},\ldots,(dB^T)_{N\!N},(dA)_{13},
\nn\\
\fl
&&
(dA)_{14},\ldots,(dA)_{1,N+\nu},(dA)_{23},\ldots,(dA)_{N,N+\nu},(dB^T)_{13},
\ldots,(dB^T)_{N-2,N}
\label{dAorder}
\eea
versus
\bea
\fl &&
(d\La_A)_{11},(d\La_A)_{12},(d\La_A)_{21},(d\La_A)_{22},(d\La_A)_{33},
.\,.\,,(d\La_A)_{N\!N},
(d\La_B)_{11},.\,.\,,(d\La_B)_{N\!N},
(d\Delta_A)_{13},
\nn\\
\fl
&&(d\Delta_A)_{14},\ldots,(d\Delta_A)_{1,N+\nu},
(d\Delta_A)_{23},\ldots,(d\Delta_A)_{N,N+\nu},
(d\Delta_B)_{13},\ldots,(d\Delta_B)_{N-2,N}\ .
\label{dOorder}
\eea
The resulting sub-Jacobi
matrix is clearly the identity matrix, and the order we have picked
is arbitrary as long as we pair $(dA)_{ij}$ with $(d\La_A)_{ij}$ or
$(d\Delta_A)_{ij}$, and respectively for $B$.

It remains to order the matrix elements $dA$ and $dB^T$ below the block
diagonal. In order to obtain sub-blocks as in eq.\ (\ref{Ladiag}) we will
always pair $(dA)_{ij}$ with the square part of  $(dB^T)_{ij}$, with $i>j$ and
$i,j\in{1,\ldots,N}$ and finish with the rectangular part of
$(dB^T)_{ij}$, with $i=N+1,\ldots,N+\nu$. For that 
we write the following partial
differentials denoted by "$|$" from eqs.\ (\ref{dA}) and (\ref{dB}) 
\bea
(dA|)_{i>j} & \equiv & \sum_{p=1}^{p<j<i}(O_A^{T}dO_A)_{ip}(\Delta_A)_{pj} -
\sum_{q>i>j}^{N+\nu}(\Delta_A)_{iq}(O_B^{T}dO_B)_{qj} \ ,
\label{partdA}\\
(dB^T|)_{i>j} & \equiv & \sum_{p=1}^{p<j<i}(O_B^{T}dO_B)_{ip}(\Delta_B)_{pj}
- \sum_{q>i>j}^{N}(\Delta_B)_{iq}(O_A^{T}dO_A)_{qj}\ .
\label{partdB}
\eea
Here we have already used that $\La_{A,B}$ are upper triangular, and that
$i>j$ lets only the independent elements of the orthogonal differentials
appear (we have chosen the below block diagonal ones).

For the variables $\Delta_{A,B}$ below the block diagonal, in order not to
interfere with the block diagonals eq.\ (\ref{Ladiag}), we introduce the
following ordering of elements:
\bea
(i,j)&\prec& (i,p) \ \mbox{if} \ p<j\ ,\nn\\
(i,j)&\prec& (q,j) \ \mbox{if} \ q>i\ .
\label{order}
\eea
Here $\prec$ implies that the matrix element $(i,j)$ on the left, that is
$(dA)_{ij}$ ($(O_A^{T}dO_A)_{ij}$) has to appear before the one on the
right \footnote{Our ordering is different from appendix A.37 in \cite{Mehta}
  for the real Ginibre ensemble.}. Looking back
to eqs.\ (\ref{partdA})  and (\ref{partdB})
this implies that the elements $(dA)_{ij}$ and
$(dB^T)_{ij}$ that depend on the most elements of $(O_A^{T}dO_A)_{ij}$ and
$(O_B^{T}dO_B)_{ij}$ will appear first, leading to a lower triangular
structure. However, the ordering eq.\ (\ref{order}) is not unique. We will
proceed $2\times2$ block-wise (plus $1\times2$ blocks for the last row when
$N$ is odd) going down the diagonal, in order to preserve the block-diagonal
structure when the matrices $\La_{A,B}$ are not diagonal.

We thus continue the labelling in eqs.\ (\ref{dAorder}) and (\ref{dOorder}) as
\bea
\fl
&&(dA)_{32},(dB^T)_{32},(dA)_{31},(dB^T)_{31},(dA)_{42},(dB^T)_{42},
(dA)_{41},(dB^T)_{41},(dA)_{54},(dB^T)_{54},\ldots,
\nn\\
\fl
&&
(dA)_{N1},(dB^T)_{N1}
\label{dAorder2}
\eea
versus
\bea
\fl &&
(O_A^{T}dO_A)_{32},(O_B^{T}dO_B)_{32},
(O_A^{T}dO_A)_{31},(O_B^{T}dO_B)_{31},
(O_A^{T}dO_A)_{42},(O_B^{T}dO_B)_{42},
(O_A^{T}dO_A)_{41},
\nn\\
\fl
&&
(O_B^{T}dO_B)_{41},
(O_A^{T}dO_A)_{54},(O_B^{T}dO_B)_{54},
\ldots,
(O_A^{T}dO_A)_{N1},(O_B^{T}dO_B)_{N1}\ .
\label{dOorder2}
\eea
It remains for us to order the rectangular part of $(dB^T)_{ij}$ with
$i=N+1,\ldots, 
N+\nu$ vs the corresponding $(O_B^{T}dO_B)_{ij}$. Using the same order as in
eq.\ (\ref{order}) we complete our Jacobi matrix by
\bea
\fl
&&(dB^T)_{N+1,N},(dB^T)_{N+1,N-1},\ldots, (dB^T)_{N+1,1},
(dB^T)_{N+2,N},\ldots,(dB^T)_{N+\nu,1}
\label{dAorder3}
\eea
versus
\bea
\fl &&
(O_B^{T}dO_B)_{N+1,N},
(O_B^{T}dO_B)_{N+1,N-1},\ldots,
(O_B^{T}dO_B)_{N+1,1},
(O_B^{T}dO_B)_{N+2,N},\ldots,\nn\\
\fl&&(O_B^{T}dO_B)_{N+\nu,1}\ .
\label{dOorder3}
\eea

In the second part of this appendix we will deal with the case of $\La_{A,B}$
being $2\times2$ block matrices rather than diagonal, which was omitted in
Section \ref{Ngen}.
We start with $N$ even, and first only deal with
the matrix elements $dA$ and the square part of $dB^T$ below the block
diagonal (eqs.\ (\ref{dAorder2}) and (\ref{dOorder2})).
In this case it is no longer sufficient to study one
pair of
neighbouring elements as in eq.\ (\ref{Ladiag}), but rather four pairs. This
leads to the following $8 \times 8$ matrix with the order chosen
above: In the columns we put
$(dA)_{i,j+1},(dB^T)_{i,j+1},(dA)_{ij},(dB^T)_{ij},(dA)_{i+1,j+1},
(dB^T)_{i+1,j+1}, 
(dA)_{i+1,j},(dB^T)_{i+1,j}$,
and in the rows the elements $(O_A^{T}dO_A)_{ij}$ and
$(O_B^{T}dO_B)_{ij}$ in the corresponding  order. Using eqs.\ (\ref{dA}) and
(\ref{dB}) this leads to the following sub-matrices 
down the diagonal for each odd $i$ and odd $j$:

\begin{tiny}
\bea
\fl&&
\left(\!
\begin{array}{cccccccc}
   (\Lambda_A)_{j+1,j+1} & -(\Lambda_B)_{ii}   &  (\Lambda_A)_{j+1,j}  &  0  &  0  &  -(\Lambda_B)_{i+1,i}  &  0  &  0  \\   -(\Lambda_A)_{ii}&  (\Lambda_B)_{j+1,j+1}  &  0   &  (\Lambda_B)_{j+1,j}  &  -(\Lambda_A)_{i+1,i}  &  0   &  0   & 0    \\
  (\Lambda_A)_{j,j+1} &  0   & (\Lambda_A)_{jj}     &  -(\Lambda_B)_{ii}   &  0    & 0      & 0      &   -(\Lambda_B)_{i+1,i} \\
   0 &  (\Lambda_B)_{j,j+1}  & -(\Lambda_A)_{ii}   &  (\Lambda_B)_{jj}  &  0  &  0  & -(\Lambda_A)_{i+1,i}     &  0   \\
   0& -(\Lambda_B)_{i,i+1}    & 0    &  0   &  (\Lambda_A)_{j+1,j+1}  &   -(\Lambda_B)_{i+1,i+1}  &  (\Lambda_A)_{j+1,j}  &   0  \\
   -(\Lambda_A)_{i,i+1} & 0    &  0  &  0  &  -(\Lambda_A)_{i+1,i+1}   &   (\Lambda_B)_{j+1,j+1} &  0  &   (\Lambda_B)_{j+1,j}  \\
   0 &  0   &  0   &   -(\Lambda_B)_{i,i+1}   &  (\Lambda_A)_{j,j+1}  &  0  &  (\Lambda_A)_{jj}   &  -(\Lambda_B)_{i+1,i+1}  \\
   0 &  0  &  -(\Lambda_A)_{i,i+1}  &  0  &   0  &  (\Lambda_B)_{j,j+1}  & -(\Lambda_A)_{i+1,i+1}   &   (\Lambda_B)_{jj} \\
\end{array}
\! \right)\nn
\eea
\end{tiny}
\be
\label{block8}
\ee
We can easily verify (using the symbolic manipulation capabilities of
Mathematica \cite{Mathematica}, for example) that the modulus of the
determinant of this is identical to 
\be\label{mini_Jacob}
{\cal J}_{ij} = \Big|(D_i-D_j)^2 + (S_i - S_j)(S_i D_j - S_j D_i)\Big|
\ee
in which $D_i$ is the determinant, and $S_i$ the trace, of the $2 \times 2$
matrix $U_i$ given by (only relevant for odd $i$)
\be
U_i \equiv
\left( \begin{array}{cc}
  (\Lambda_A)_{ii} & (\Lambda_A)_{i,i+1} \\
  (\Lambda_A)_{i+1,i} & (\Lambda_A)_{i+1,i+1}
\end{array} \right)
\left( \begin{array}{cc}
  (\Lambda_B)_{ii} & (\Lambda_B)_{i,i+1} \\
  (\Lambda_B)_{i+1,i} & (\Lambda_B)_{i+1,i+1}
\end{array} \right).
\ee
The same definition obviously applies for subscript $j$.
However, $D_i$ and $S_i$ can of course be written in terms of the eigenvalues
of $U_i$, which we denoted $\Lambda_i^2$ and $\Lambda_{i+1}^2$, i.e.
\be
D_i  = \det U_i= \Lambda_i^2 \Lambda_{i+1}^2\ , \ \
S_i =  \Tr\, U_i= \Lambda_i^2 + \Lambda_{i+1}^2.
\ee
Substituting these into eq.\ (\ref{mini_Jacob}), and factorising, gives
\be
{\cal J}_{ij} =
\Big|(\Lambda_i^2-\Lambda_j^2)(\Lambda_{i+1}^2-\Lambda_j^2)
(\Lambda_i^2-\Lambda_{j+1}^2)(\Lambda_{i+1}^2-\Lambda_{j+1}^2)\Big|.
\ee
This is exactly what we got in the case when the $\Lambda_A$ and $\Lambda_B$
matrices were diagonal, i.e. the product of four copies of
eq.\ (\ref{Ladiag}), one for each of the combinations $(i,j)$, $(i+1,j)$,
$(i,j+1)$ and $(i+1,j+1)$.

When $N$ is odd, we can treat the $2 \times 2$ blocks up to $N-1$ (inclusive)
as before. We still have to consider the final column of $dA$, etc., and it is
necessary here to treat $(j,N)$ and $(j+1,N)$ together as $2\times1$ blocks
(for odd values of $j$ in the range $1 \leq j < N$).
Hence, we have extra diagonal blocks in the
Jacobian (for each odd $j$) of the form
\be
\left(
\begin{array}{cccc}
 (\Lambda_A)_{j+1,j+1} & -(\Lambda_B)_{NN} & (\Lambda_A)_{j+1,j} & 0  \\
 -(\Lambda_A)_{NN}   &  -(\Lambda_B)_{j+1,j+1}&0    & (\Lambda_B)_{j+1,j}\\
 (\Lambda_A)_{j,j+1}& 0 & (\Lambda_A)_{jj} & -(\Lambda_B)_{NN}\\
 0 & (\Lambda_B)_{j,j+1} & -(\Lambda_A)_{NN} & (\Lambda_B)_{jj}\\
\end{array}
\right)
\ee
which is in fact the first sub-block of eq.\ (\ref{block8}) with $i=N$.
The modulus of the determinant of this is identically equal to
\be
{\cal J}_{Nj} = \Big| D_j - S_j \Lambda_N^2 + \Lambda_N^4 \Big|,
\ee
where we used that $(\Lambda_A)_{NN}(\Lambda_B)_{NN} = \Lambda_N^2$ (no sums),
and $D_j$ and $S_j$ were defined as before. But we can switch to writing $D_j$
and $S_j$ in terms of the eigenvalues of $U_j$, and this gives
\be
{\cal J}_{Nj} = \Big|(\Lambda_N^2-\Lambda_j^2)
(\Lambda_N^2-\Lambda_{j+1}^2)\Big|.
\ee
We see that this is again of the expected form.

Let us turn to the remaining variables, the rectangular part of $dB^T$ in
eqs.\ (\ref{dAorder3}) and (\ref{dOorder3}). We first suppose that $N$ is even.
For each $i$ and each odd $j$ we find that the variables
$(dB^T)_{i,j+1},(dB^T)_{ij}$ and
$(O_B^{T}dO_B)_{i,j+1},(O_B^{T}dO_B)_{ij}$ are now coupled into the following
$2\times2$ blocks:
\be
\left(
\begin{array}{cc}
(\Lambda_B)_{j+1,j+1} & (\Lambda_B)_{j+1,j} \\
(\Lambda_B)_{j,j+1}   & (\Lambda_B)_{jj}
\end{array}
\right)
\ee
whose determinant is (after a relabelling $2j-1 \rightarrow j$) simply
$\det B_j$ (using the definition of $B_j$ after eq.\ (\ref{2nd_prod_result})).
This is independent of $i$ which takes $\nu$ different values,
$N+1 \le i \le N+\nu$.
We therefore have a total contribution to the Jacobian of
\be
\prod_{j=1}^{N/2} |\det B_j|^{\nu}\ \mbox{for $N$ even,}
\ee
which is exactly as before (eq.\ (\ref{2nd_prod_result})).

Finally, when $N$ is odd, we just pick up an extra factor of
$(\Lambda_B)_{NN}$ from each of the $\nu$ cells in the last column
$(dB^T)_{iN}$, and so
\be
\prod_{j=1}^{(N-1)/2} |\det B_j|^{\nu} |(\Lambda_B)_{NN}|^\nu
\ \mbox{for $N$ odd.}
\ee
This ends the calculation for non-diagonal matrices $\La_{A,B}$.

\end{appendix}



\section*{References}
\begin{thebibliography}{99}


\bibitem{Ginibre65} J. Ginibre, J. Math. Phys. {\bf 6}, 440 (1965).

\bibitem{FS} Y.V. Fyodorov and H.-J. Sommers,
{ J. Phys. {\bf A}: Math. Gen.} {\bf 36} (2003) 3303\newline
[arXiv:nlin.CD/0207051].

\bibitem{Lehmann91} N. Lehmann and H.-J. Sommers, Phys. Rev. Lett. {\bf 67},
  941 (1991).

\bibitem{Edelman97} A. Edelman, J. Multivariate Anal. {\bf 60}, 203 (1997).

\bibitem{Akemann07}
 E. Kanzieper and G. Akemann, Phys. Rev. Lett. {\bf 95},
  230201 (2005) [arXiv:math-ph/0507058];\newline
G. Akemann and E. Kanzieper, J. Stat. Phys. {\bf 129},
  1159 (2007) [arXiv:math-ph/0703019].

\bibitem{Sinclair06} C.D. Sinclair,
Int. Math. Res. Not. {\bf 2007} rnm015 (2007)
[arXiv:math-ph/0605006].

\bibitem{Sommers2007} H.-J. Sommers,   J. Phys. {\bf A40}, F671  (2007)
[arXiv:0706.1671].

\bibitem{Forrester07} P.J. Forrester and T. Nagao, Phys. Rev. Lett. {\bf 99}
  050603 (2007) [arXiv:0706.2020 [cond-mat.stat-mech]];
J. Phys.  {\bf A41},
375003 (2008) [arXiv:0806.0055 [math-ph]]

\bibitem{Borodin07} A. Borodin and C.D. Sinclair, arXiv:0706.2670v2 [math-ph];
arXiv:0805.2986 [math-ph].

\bibitem{Sommers2008}
H.-J. Sommers and W. Wieczorek,  J. Phys.  {\bf A41},
405003 (2008) [arXiv:0806.2756 [cond-mat.stat-mech]].

\bibitem{ForresterMays}
P.J. Forrester and A. Mays, 	arXiv:0809.5116v2 [math-ph]

\bibitem{SV93}
E.V. Shuryak and J.J.M. Verbaarschot, Nucl. Phys. {\bf A560} (1993) 306
[arXiv:hep-th/9212088];\newline
J. Verbaarschot, Phys. Rev. Lett. {\bf 72} (1994) 2531 
[arXiv:hep-th/9401059v1].

\bibitem{Steph} M.A. Stephanov, { Phys. Rev. Lett.} {\bf 76} (1996) 4472.
[arXiv:hep-lat/9604003].

\bibitem{HOV} M.A. Halasz, J.C. Osborn, and J.J.M. Verbaarschot,
Phys. Rev. {\bf D56} (1997) 7059\newline
[arXiv:hep-lat/9704007].

\bibitem{KJreplica} K. Splittorff and J.J.M. Verbaarschot,
Nucl. Phys. {\bf B683} (2004) 467 
[arXiv:hep-th/0310271v3]

\bibitem{Osborn} J.C. Osborn,
Phys. Rev. Lett. {\bf 93} (2004) 222001
[arXiv:hep-th/0403131].

\bibitem{A03} G. Akemann,
J. Phys. {\bf A}: Math. Gen. {\bf 36} (2003) 3363 [arXiv:hep-th/0204246].

\bibitem{AOSV}
G. Akemann, J.C. Osborn, K. Splittorff, and J.J.M. Verbaarschot,
Nucl. Phys. {\bf B712} (2005) 287.
[arXiv:hep-th/0411030].

\bibitem{AB}
G. Akemann,
  Nucl.\ Phys. {\bf B730}, 253 (2005)   [arXiv:hep-th/0507156];
G. Akemann and F. Basile,
Nucl. Phys. {\bf B766}, 150 (2007)
[arXiv:math-ph/0606060].

\bibitem{APSo}
G. Akemann, M.J. Phillips, and H.-J. Sommers,
J. Phys. {\bf A}: Math. Theor. {\bf 42} (2009) 012001
[arXiv:0810.1458v1 [math-ph]].

\bibitem{A07mu} J.J.M. Verbaarschot,
Les Houches Summer School, France, 6-25 June 2004,
arXiv:hep-th/0502029v1;\newline
G. Akemann,
Int. J. Mod. Phys. {\bf A22} (2007) 1077 [arXiv:hep-th/0701175].

\bibitem{Magnea08}
D. Bernard and A. LeClair,
{\it A Classification of Non-Hermitian Random Matrices},
proceedings of the NATO Advanced Research Workshop on Statistical Field
Theories, Como 18-23 June 2001, [arXiv:cond-mat/0110649];\newline
U. Magnea,
J. Phys.  {\bf A41}, 045203 (2008) [arXiv:0707.0418v2 [math-ph]].

\bibitem{FM09}
P.J. Forrester and A. Mays, arXiv:0910.2531 [math-ph]

\bibitem{EKS94} A. Edelman, E. Kostlan, and M. Shub, J. Amer. Math. Soc.
{\bf  7}, 247 (1994).

\bibitem{AV}
G. Akemann and G. Vernizzi,
  Nucl.\ Phys. {\bf B660}, 532 (2003) [arXiv:hep-th/0212051].

\bibitem{TW}
C.A. Tracy and H. Widom,
J. Stat. Phys. {\bf 92} (1998) 809.

\bibitem{F99}
P.J. Forrester, T. Nagao, and G. Honner,
Nucl. Phys. {\bf B553} (1999) 601 
[arXiv:cond-mat/9811142v1 [cond-mat.mes-hall]].

\bibitem{Efetov97} K.B. Efetov, Phys. Rev. Lett. {\bf 79} 491 (1997)
[arXiv:cond-mat/9702091 [cond-mat.dis-nn]].

\bibitem{Sommers88} H.-J. Sommers, A. Crisanti, H. Sompolinsky, and Y. Stein,
  Phys. Rev. Lett {\bf 60} (1988) 1895.

\bibitem{FKS98} Y.V. Fyodorov, B.A. Khoruzhenko, and H.-J. Sommers,
Ann. Inst. Henri Poincar\'e
{\bf 68}, 449 (1998) [arXiv:chao-dyn/9802025].

\bibitem{Mehta}
M.L. Mehta, {\it Random Matrices}, Academic Press, Third
Edition, London (2004).

\bibitem{FKS} Y.V. Fyodorov, B.A. Khoruzhenko, and H.-J. Sommers,
Phys. Lett. {\bf A226} (1997) 46  [arXiv:cond-mat/9606173];
Phys. Rev. Lett. {\bf 79} (1997) 557  [arXiv:cond-mat/9703152].

\bibitem{Abramowitz} M. Abramowitz and I.E. Stegun, {\it Handbook of
Mathematical Functions}, Dover Publications Inc., New York (1965).

\bibitem{Grad} I.S. Gradshteyn and I.M. Ryzhik, {\it Table of Integrals,
Series and Products}, 6th Edition, Academic
  Press, London (2000).

\bibitem{JacNc2}
J. Verbaarschot,
Nucl. Phys. {\bf B426} (1994) 559 
[arXiv:hep-th/9401092v1].

\bibitem{Mathematica} Wolfram Research, Inc., {\it Mathematica},
Version 7.0, Champaign, IL (2008)


\end{thebibliography}
\end{document}